\documentclass[aps,onecolumn,balancelastpage,amsmath,amssymb,nofootinbib,preprint,12pt,]{revtex4-1}

\usepackage[T1]{fontenc} 

\usepackage{yfonts}

\usepackage[utf8]{inputenc}
\usepackage{tikz}
\usepackage{capt-of}

\usepackage{comment}

\usepackage{xcolor}
\definecolor{darkblue}{rgb}{0.15,0.35,0.55}
\definecolor{reddish}{rgb}{0.65, 0.2, 0.2}
\definecolor{purplish}{rgb}{.2, .2, .5}

\usepackage{color}   
\usepackage{hyperref}
\hypersetup{
    colorlinks=true, 
    linktoc=black,     
    linkcolor=blue,  
    citecolor=red
}

\addtocontents{toc}{\protect\enlargethispage{\baselineskip}}

\usepackage{calrsfs}
\usepackage{tikz}

\usepackage[rightcaption]{sidecap}
\usepackage{graphicx}
\graphicspath{ {./images/} }

\newcommand{\m}{\mu}
\newcommand{\n}{\nu}

\newcommand{\vev}[1]{{\left< {#1} \right>}}

\newcommand{\cur}[1]{{\left( {#1} \right)}}
\newcommand{\squ}[1]{{\left[ {#1} \right]}}

\newcommand{\tr}{{\rm tr\,}}

\DeclareMathAlphabet{\pazocal}{OMS}{zplm}{m}{n}

\newcommand{\cA}{{\pazocal A}}
\newcommand{\cB}{{\pazocal B}}

\newcommand{\cD}{{\pazocal D}}

\newcommand{\cP}{{\pazocal P}}

\newcommand{\cM}{{\pazocal M}}

\newcommand{\cF}{{\pazocal F}}
\newcommand{\cG}{{\pazocal G}}
\newcommand{\cV}{{\pazocal V}}

\newcommand{\cH}{{\pazocal H}}

\newcommand{\bK}{{\mathcal K}}

\newcommand{\bP}{{\mathcal P}}
\newcommand{\bQ}{{\mathcal Q}}

\newcommand{\bF}{{\mathcal F}}
\newcommand{\bG}{{\mathcal G}}

\newcommand{\bR}{{\mathcal R}}

\newcommand{\bJ}{{\mathcal J}}

\usepackage{blindtext}

\usepackage{lipsum}

\begin{document}

\title{\boldmath Higher-form and (non-)Stückelberg symmetries \\ in non-equilibrium systems\\ \bigskip}
\author{Michael J. Landry} 
\affiliation{Department of Physics, Center for Theoretical Physics,\\ Columbia University, 538W 120th Street, New York, NY, 10027, USA \bigskip }

\begin{abstract} \bigskip We investigate the role of higher-form symmetries in non-equilibrium systems from the perspective of effective actions defined on the Schwinger-Keldysh contour. To aid our investigation, we extend the coset construction to account for $p$-form symmetries at zero and finite temperature. Additionally we investigate how, out of equilibrium, symmetries of the action need not lead to meaningful conserved currents at the level of the equations of motion. For reasons that will become apparent, we term symmetries with conserved currents Stückelberg symmetries and those without meaningful conserved currents non-Stückelberg symmetries (NSS).
Ordinarily any action constructed exclusively from building-blocks furnished by the coset construction will have Stückelberg symmetries associated with each symmetry generator. To expand the set of systems describable by the coset construction, we devise a method by which NSS generators can be included as well. While 0-form NSS are quite common in non-equilibrium effective actions, the introduction of $p$-form NSS is novel. We use these $p$-form NSS to investigate spontaneous symmetry breaking of $p$-form symmetries. 
We find that in non-equilibrium systems, whether or not a symmetry appears spontaneously broken can depend on the time-scale over which the system is observed. Finally, using our new coset construction, we formulate actions for a number of systems including chemically reacting fluids, Yang-Mills theory, Chern-Simons theory, magnetohydrodynamic systems, and dual superfluid and solid theories.

\end{abstract}

\maketitle

\tableofcontents

\section{Introduction}

The physics of many-body systems out of finite-temperature equilibrium is a broad area of study with applications to almost all aspects of physics. Such systems typically resist any attempt at an exact description; they are simply too complicated and chaotic. If, however, we are interested only in the extreme infrared (IR) then we can use various approximation schemes that make the descriptions of non-equilibrium many-body systems tractable. In particular, it is often possible to construct an effective field theory (EFT) consisting of a small number of IR degrees of freedom. The primary principle guiding the construction of such EFTs is symmetry; in particular, once the field content is specified, an effective action can be constructed by writing down a linear combination of all symmetry-invariant terms at a given order in a derivative expansion~\cite{Nicolis,Zoology,More gapped Goldstones,coset,Finite T superfluid}. In fact, Landau's classification {\it defines} states of matter according to their spontaneous symmetry breaking (SSB) pattern. 

Ordinary actions necessarily describe conservative systems, meaning they cannot adequately describe finite-temperature systems, which are inherently dissipative. However, recent work has extended the action principle by employing the in-in formalism, which enables the construction of a non-equilibrium effective action defined on the Schwinger-Kedysh (SK) contour~\cite{Grozdanov:2013dba,H. Liu,H. Liu 2,H. Liu 3,H. Liu 2.2,H. Liu 2.3,H. Liu 2.1,H. Liu 4,Harder:2015nxa,Constraints on Fluid,Jensen:2017kzi,Jensen:2018hse,Jensen:2018hhx,Towards hydrodynamics,Kovtun:2014hpa,Haehl:2017zac,Haehl:2018lcu,Fluid manifesto,Hongo 1,Hongo 2,Hongo 3,Hongo 4,Volkov,Landry,Landry chemistry,Landry quasicrystal,Landry second sound}. Such non-equilibrium effective actions account for dissipation and statistical fluctuations arising from thermal and quantum effects. The price we pay for dissipation is a doubling of the field content and an action with complex-valued coefficients. 

While ordinary symmetries provide a powerful guiding principle for constructing EFTs, they are not the whole story. This is particularly true for non-equilibrium systems. In this paper we focus on two main extensions of ordinary symmetry:
\begin{itemize}
\item There are many phases of matter that are characterized by topological features, known as {\it topological phases of matter}. It was long believed that these topological phases could not be classified according to symmetry principles alone; however, recent work has challenged this idea~\cite{Gaiotto:2014kfa}. In particular, it is suggested that a kind of generalized symmetry principle---known as {\it higher-form symmetries}---can be employed to describe topological features. In this way, topological properties of many-body systems can be classified according to Landau's approach. The Noether charges corresponding to these higher-form symmetries describe the conservation of higher-dimensional extended objects. For example a $p$-form charge counts the number of charged $p$-dimensional objects. Such higher-form symmetries arise in many areas including gauge theories, magnetohydrodynamics, dual theories of superfluids and solids, and many others~\cite{Gaiotto:2014kfa,Zhao:2020vdn,Lake:2018dqm,Hidaka:2020ucc,Hofman:2018lfz,Delacretaz:2019brr,Harlow:2018tng,Grozdanov:2016tdf,Grozdanov:2018ewh,MHD,Grozdanov:2018fic,Grozdanov:2017kyl}. In fact, a common feature of systems with spontaneously broken $U(1)$ symmetries---like superfluids and solids~\cite{Landry}---is the existence of an emergent higher-form symmetry with mixed 't Hooft anomaly~\cite{Iqbal:2020lrt}. Further, when a higher-form symmetry is spontaneously broken, a generalized Goldstone theorem guarantees the existence of a gapless mode.

\item The role of symmetries is somewhat different for non-equilibrium systems than for systems at zero-temperature that are described by ordinary actions. In particular, for ordinary actions, Noether's theorem furnishes a one-to-one correspondence between physical symmetries and conserved quantities. In non-equilibrium effective actions, however, no such one-to-one correspondence is guaranteed. To ensure that a given symmetry enjoys a corresponding conserved current, it is helpful to introduce external gauge fields for that symmetry. Then the low-energy fluctuations are described by dynamical Stückelberg fields. Therefore we refer to symmetries with conserved currents as Stückelberg symmetries and those without conserved current as non-Stückelberg symmetries (NSS).

\end{itemize} 

It is the aim of this paper to systematically investigate how higher-form symmetries and NSS can be used to construct non-equilibrium effective actions. To aid in the systematization of our approach, we employ the coset construction, which is a powerful tool that facilitates the formulation of effective actions for Goldstone and Goldstone-like excitations~\cite{Landry,Ogievetsky,Ivanov and Ogievetsky,Low,Weinberg,Wheel,Landry spin}. The basic idea behind the coset construction is that the Maurer-Cartan one-form, parameterized by Goldstone fields, furnishes symmetry-invariant building-blocks, which are then used to formulate an effective action. We find that for spontaneously broken $p$-form symmetries, the corresponding equivalent of the Maurer-Cartan form is no longer a one-form but is instead a $p+1$-form. When an ordinary or higher-form symmetry is spontaneously broken, a (generalized) Goldstone theorem guarantees the existence of a gapless mode~\cite{Zhao:2020vdn}. For non-equilibrium systems, even unbroken symmetries can have corresponding gapless modes that resemble Goldstone excitations. At the level of the coset construction, we parameterize the {\it full} symmetry group with dynamical Goldstone fields. Then to distinguish between the true Goldstones corresponding broken symmetries from the Goldstone-like fields corresponding to unbroken symmetries, we endow the unbroken Goldstones with time-independent, right-acting gauge redundancies~\cite{Landry}. We extend these gauge-redundancies to account for gapless modes corresponding to unbroken $p$-form symmetries. Although the modes corresponding to unbroken symmetries are technically not true Goldstones, we will nevertheless refer to all fields that appear in the coset parameterization as Goldstones. 
Further, we find that any action constructed exclusively using terms furnished by the Maurer-Cartan form will have a one-to-one correspondence between physical symmetries and conserved quantities---that is all symmetry generators in the coset construction correspond to Stückelberg symmetries. In order to extend the coset construction to describe more general phenomena, we identify a procedure by which NSS charges can be included.

To build the readers's intuition for higher-form symmetries and NSS, we investigate the non-equilibrium properties of finite-temperature electromagnetism in various settings at the level of quadratic effective actions. We use  these various examples to illustrate key features of non-equilibrium systems: (1) whether or not a symmetry appears spontaneously broken depends on the time-scale over which the system is observed,  (2) the currents associated with $p$- and $d-p-1$-form symmetries with mixed anomaly have a definite mathematical relationship, and (3) there is close connection between mixed 't Hooft anomalies, NSS, SSB, and propagative versus diffusive dispersion relations.  \\

Throughout this work, we use the `mostly plus' convention for the Minkowski spacetime metric $\eta_{\mu\nu} = \text{diag}(-,+,\dots,+)$.

\section{Higher-form symmetries: a review}

Suppose a relativistic quantum field theory in $3+1$ dimensions has conserved $U(1)$ current 
\begin{equation}J^\mu,~~~~~~~~~~ \partial_\mu J^\mu = 0.\end{equation}
The usual way to define a conserved charge is by integrating the charge density over the volume of space, that is 
\begin{equation}\label{original charge}Q = \int d^3 x \,J^0,~~~~~~~~~~ \frac{d}{dt} Q = 0, \end{equation}
but there is another perspective that involves differential forms. In particular, consider the Hodge star dual of the current
\begin{equation}\star J_{\mu\nu\lambda} = \epsilon_{\mu\nu\lambda\rho} J^\rho, \end{equation}
where $\epsilon_{\mu\nu\lambda\rho}$ is the Levi-Civita tensor. Notice that $\star J$ is a three-form, which can naturally be integrated over a three-dimensional hyper-surface in spacetime. Let $\Sigma$ be such a hyper-surface. Then we can define the charge as a function of the hyper-surface by
\begin{equation} Q[\Sigma] = \int_\Sigma \star J .  \end{equation} 
If we take $\Sigma$ to be a spatial volume at fixed time, then we recover the original expression for the charge~(\ref{original charge}). Thus the differential form definition of charge contains within it the standard formulation. Notice that the conservation of $J$ implies that $d\star J=0$, that is, $\star J$ is closed. Therefore, if we continuously deform the hyper-surface $\Sigma\to \Sigma'$ in such a way that it leaves the boundary unchanged and never intersects any charged operator insertions, then $Q[\Sigma] = Q[\Sigma']$. In this sense, the charge's dependence on $\Sigma$ is purely topological. Moreover, this topological dependence can be interpreted as a conservation law in the sense that we can translate $\Sigma$ through time without changing the total charge as long as we do not encounter any sources (i.e. charged operator insertions).\footnote{Translating $\Sigma$ through time will generally change the boundary; however, if $\Sigma$ is a constant time slice, then it extends to spatial infinity and hence has no boundary.} Finally, we should comment on the dimensionality of the objects that are conserved. The charge $Q[\Sigma]$ counts the total number of point-particles on the slice $\Sigma$. From another perspective, however, point-particles travel through spacetime along their respective worldlines, so they ought to be thought of as one-dimensional objects (or $D_0$-branes). Thus the charge $Q[\Sigma]$ counts the number of worldlines that intersect the hyper-surface $\Sigma$. 

At this point, we can generalize the notion of current and charge. In the interest of complete generality, 
we will now work in a $d+1$ dimensional spacetime. Suppose a current with $p+1$ indices 
\begin{equation} \bJ^{\mu_0\dots\mu_{p}},~~~~~\partial_{\mu_0} \bJ^{\mu_0\dots\mu_{p}} =0  ,\end{equation}
that is antisymmetric under interchange of any adjacent pair of indices. 
Then the Hodge star dual~is
\begin{equation}\star \bJ_{\mu_1\dots\mu_{d-p}} = \epsilon_{\mu_1\dots\mu_{d-p} \nu_0\dots\nu_{p} } J^{\nu_0\dots\nu_p}.  \end{equation} 
Since $\star \bJ$ is a $d-p$-form, it is naturally integrated over an $d-p$-dimensional hyper-surface $\Sigma_{d-p}$. We therefore define the charge by
\begin{equation}\bQ[\Sigma_{d-p}] = \int _{\Sigma_{d-p}} \star \bJ. \end{equation}
Notice that $d\star\bJ=0$, meaning that the dependence of $\bQ$ on the choice of hyper-surface $\Sigma_{d-p}$ is topological, which in turn implies that $\bQ$ is conserved. Finally, since we are integrating $\star \bJ$ over a surface of dimension $d-p$, the dimension of the conserved objects will not in general be point-particles traveling along worldlines (unless $p=0$). Instead, $\bQ[\Sigma_{d-p}]$ counts the number of charged $p+1$-dimension objects---or $D_p$-branes---that intersect $\Sigma_{d-p}$.  We say that a system with conserved current $\bJ$ has an $p$-form symmetry. Thus, in this language, ordinary symmetries are zero-form symmetries.  \\

We have made reference to charged operator insertions, but so far have not identified what they are. The objects charged under a $p$-form symmetry must themselves be $p$-dimensional objects that cannot be composed of lower-dimensional operators. Suppose we have a $p$-form field $\varphi^p$ that enjoys some sort of gauge symmetry. For the sake of concreteness, suppose that we have the gauge symmetry $\varphi^p\to \varphi^p+d\gamma$ for $p-1$-form $\gamma$. Then, given a closed $p$-dimensional manifold $C_p$, we may construct the $p$-dimensional generalization of a Wilson loop by
\begin{equation} W[C_p] = e^{i\int_{C_p} \varphi^p } .  \end{equation}
The charge $\bQ[\Sigma_{d-p}]$ can detect the presence of the $p$-dimensional Wilson surfaces if and only if $\Sigma_{d-p}$ and $C_p$ are linked. In this way, $\bQ[\Sigma_{d-p}]$ counts the number of closed $p$-dimensional Wilson surfaces that link with $\Sigma_{d-p}$.

\subsection{Superfluids at zero temperature}\label{Superfluids at zero temperature}

A relativistic superfluid is a system that, in addition to Poincaré symmetry, enjoys a spontaneously broken $U(1)$ internal symmetry $N$, which corresponds to particle number conservation. Further, superfluids exist at finite charge density. We assume that in the deep infrared, the only gapless mode is the Goldstone $\psi$ associated with the broken charge $N$. The leading order-action is therefore
\begin{equation}\label{superfluid action T=0} S = \int d^4 x P(X),~~~~~ X = \sqrt{-\partial_\mu \psi \partial^\mu\psi}, \end{equation} 
and the current corresponding to $N$ is 
\begin{equation}\label{U(1) current superfluids} J_{U(1)} ^\mu = - \frac{P'(X)}{X} \partial^\mu\psi.  \end{equation} 
Notice that in equilibrium, since charge density is non-zero, $J_{U(1)} ^\mu\propto \delta^\mu_0$, meaning that $\partial_\mu\psi \propto \delta_\mu^0$. Thus, $\psi$ has a time-dependent background $\vev{\psi}
 = \mu_0 t$. It turns out that $\mu_0$ can be interpreted as the equilibrium chemical potential. 
 
By virtue of possessing a $U(1)$ Goldstone, superfluids also enjoy a two-form symmetry with corresponding current  
\begin{equation} K^{\mu\nu\lambda} = \epsilon^{\mu\nu\lambda\rho} \partial_\rho \psi.   \end{equation} 
Notice that $\partial_\mu K^{\mu\nu\lambda} = 0$ identically as partial derivatives commute. The fact that $K$ is conserved as a mathematical identity may give some readers pause: is there any physical content to this conservation equation? It turns out the answer is emphatically, ``yes!'' To see how this can be, we will recast the theory of superfluids in terms of a dual theory involving a two-form gauge field $A_{\mu\nu}$. 

Begin by replacing $\partial_\mu \psi \to V_\mu$, for some arbitrary one-form $V$. Then we can define an auxiliary action
\begin{equation}\label{aux S superfluid} S_\text{AUX} = \int d^4 x \cur{P(\cV) -\frac{1}{2} \epsilon^{\mu\nu\lambda\rho} V_\mu \partial_\nu A_{\lambda\rho}}, \end{equation} 
where $\cV$ is the magnitude of $V_\mu$. Notice that the equations of motion for $A_{\mu\nu}$ are 
\begin{equation}\partial_{[\mu}V_{\nu]}=0,\end{equation}
which, if spacetime is topologically trivial, implies 
$ V_\mu = \partial_\mu \psi$ for some scalar $\psi$. Plugging this solution back into $S_\text{AUX}$, we recover the original superfluid action~(\ref{superfluid action T=0}). Alternatively, we can integrate out $V_\mu$, in which case we find the equations of motion
\begin{equation}\label{eq for V} -\frac{P'(\cV)}{\cV} V^\mu = F^\mu,\end{equation}
where we have defined the field strength by
\begin{equation}F^\mu = \frac{1}{2}\epsilon^{\mu\nu\lambda\rho} \partial_\nu A_{\lambda\rho}. \end{equation}
We can then algebraically solve~(\ref{eq for V}) to find $V$ in terms of $F$.\footnote{In general the equation to be solved may be very complicated, but all we require is that a solution exist.} Plugging this result back into~(\ref{aux S superfluid}), we obtain a dual action that only depends on $Y=\sqrt{-F^\mu F_\mu}$. Explicitly, we have
\begin{equation} S_\text{DUAL} = \int d^4 x \, L(Y), \end{equation}
for some function $L$, where $L(Y) \equiv P(\cV(Y))$. Notice that this action is invariant under the gauge symmetry
\begin{equation} A_{\mu\nu} \to A_{\mu\nu} + \partial_{[\mu} \lambda_{\nu]}(x).\end{equation}

The equations of motion of the dual action are
\begin{equation} \partial_\mu \frac{\partial S_\text{DUAL}}{\partial (dA)_{\mu\nu\lambda}} = 0.  \end{equation}
But this equation is just the conservation law for a higher-form current. We therefore identify
\begin{equation} K^{\mu\nu\lambda } =  \frac{\partial S_\text{DUAL}}{\partial (dA)_{\mu\nu\lambda}}. \end{equation}
We now see that the conservation of $K$ is no longer a mere mathematical identity, but is a manifestly physical equation describing the dynamics of superfluids. But what has happened to $J_{U(1)}^\mu$? Comparing equations~(\ref{U(1) current superfluids}) and~(\ref{eq for V}), we have the identification
\begin{equation}J_{U(1)}^\mu = F^\mu.\end{equation}
But notice that $\partial_\mu F^\mu=0$ identically owing to the fact that partial derivatives commute. Thus, in the dual picture the particle-number conservation equation is a mere mathematical identity. 

\subsection{Electromagnetism}

We now consider the theory of electromagnetism as a system involving higher-form symmetries. Letting $A$ be the photon field and $F=dA$ be the field-strength tensor, the action for free Maxwell theory in 3+1 dimensions is
\begin{equation}S = -\frac{1}{4}\int d^4 x \, F^2. \end{equation}
The equations of motion are $\partial_\mu F^{\mu\nu} = 0$, meaning that we have a one-form symmetry with corresponding conserved current $J_\text{el}^{\mu\nu}\equiv F^{\mu\nu}$. Then given a closed two-dimensional surface $\Sigma_2$, we have the conserved charge
\begin{equation}\bQ[\Sigma_2] = \int _{\Sigma_2} \star F.  \end{equation}
Taking $\Sigma_2$ to exist at constant time, we therefore have
\begin{equation}\bQ[\Sigma_2] = \int _{\Sigma_2} \hat n\cdot \vec E,\end{equation}
where $\hat n$ is normal to the surface $\Sigma_2$ and $E^i =F^{0i}$. We thus recognize this one-form charge as counting the number of electric field lines passing through the Gaussian surface $\Sigma_2$. 

What are the charged objects? As we mentioned earlier, they are Wilson loops. For closed curve $C$, we have
\begin{equation}W[C]=e^{i\int_C A} . \end{equation}
Physically, $C$ corresponds to the worldline of a massive charged particle, which is a source for the electric field lines. In this way, if $C$ links with $\Sigma_2$, then the amount of electric flux through $\Sigma_2$ changes according to Gauss's law. We can see, therefore, that $\bQ[\Sigma_2]$ counts the number of Wilson loops that link with $\Sigma_2$. 

Electromagnetism has another one-form symmetry with corresponding current $J_\text{mag}^{\mu\nu} = \star F^{\mu\nu}$, which is conserved as an identity. This charge counts the number of magnetic field lines passing through a given Gaussian surface. When dealing with the full, sourced Maxwell theory, $J_\text{el}^{\mu\nu}$ is no longer exactly conserved, whereas, in the absence of magnetic monopoles, $J_\text{mag}^{\mu\nu}$ is always exactly conserved. As a result, when we consider magnetohydrodynamics, $J_\text{mag}^{\mu\nu}$ will play a key role.

\section{Non-equilibrium EFT: a review}
In this section, we review the fundamentals of non-equilibrium EFT. We will merely highlight the key points of this formalism; for more in-depth discussions consult~\cite{H. Liu, Landry}. 

\subsection{The basics}

The goal is to construct effective actions that capture the long-distance and late-time behavior of systems out of thermal equilibrium. As a result, the equilibrium density matrix about which we will perturb is the standard thermal matrix given by
\begin{equation}\rho_0 =\frac{e^{-\beta_0 P^0}}{\text{tr} (e^{-\beta_0 P^0})},\end{equation} 
where $P_0$ is the time-translation generator and $\beta_0$ is the inverse equilibrium temperature. Because the equilibrium state is mixed (i.e. not a pure quantum state), we must use the in-in formalism and define our effective action on the Schwinger-Keldysh (SK) contour. For every quantum field $\varphi$, we have two copies corresponding to the two legs of the SK contour: $\varphi_1$ lives on the forward contour and $\varphi_2$ lives on the backward contour. We write $\varphi_s$ for $s=1,2$. It is often convenient to work in the retarded-advanced basis
\begin{equation} \varphi_r=\frac{1}{2} (\varphi_1+\varphi_2),~~~~~\varphi_a = \varphi_1-\varphi_2.  \end{equation}
It turns out that $\varphi_r$ acts as a classical field (or the expectation value of a quantum field) and $\varphi_a$  contains information about thermal and quantum fluctuations. Denoting the effective action by $I_\text{EFT}[\varphi_1,\varphi_2]$, we claim without proof that $I_\text{EFT}$ is subject to the conditions that follow.

\begin{itemize}
     \item The coefficients of $I_\text{EFT}$ are complex. There are three important constraints that come from unitarity, namely 
\begin{equation}\begin{split}\label{unitarity}I^*_\text{EFT}[\varphi_{1},\varphi_{2}]&=-I_\text{EFT}[\varphi_{2},\varphi_{1}]
\\  \text{Im} I_\text{EFT}[\varphi_{1},\varphi_{2}]&\geq 0, ~~\text{for any} ~~\varphi_{1},~\varphi_{2}
\\ I_\text{EFT}[\varphi_{1}=\varphi_{2}]&=0.
\end{split}\end{equation}
     \item Any symmetry of the UV theory is a symmetry of $I_\text{EFT}$, except for time-reversing symmetries. If the equilibrium density matrix $\rho$ takes the form of a thermal matrix, $\rho\propto e^{-\beta_0 \bar P_0}$, and the UV theory possesses some kind of anti-unitary time-reversing symmetry $\Theta$, then our EFT will enjoy the symmetries
\begin{equation}\begin{split}\label{quantum dynamical KMS} \varphi_{1} (x)&\to\Theta \varphi_{1}(t-i\theta,\vec x),
\\ \varphi_{2} (x)&\to\Theta \varphi_{2} (t+i(\beta_0-\theta),\vec x),
\end{split}\end{equation} 
for any $\theta\in[0,\beta_0]$, known as the dynamical KMS symmetries. In the classical limit, which is formally given by $\hbar\to 0$, the dynamical KMS symmetries become
\begin{equation} \begin{split} 
\varphi_r(x) &\to \Theta\varphi_r(x),\\
\varphi_a(x) & \to \Theta \varphi_a +i\Theta[\beta \partial_t \varphi_r]. 
\end{split}\end{equation}
Because the change in $\varphi_a$ involves a temporal derivative of $\varphi_r$, we consider $\varphi_a$ and $\beta_0\partial_t\varphi_r$ to contribute at the same order in the EFT derivative expansion. 
\item In the distant future, the fields on the 1 and 2 legs of the SK contour must coincide, that is $\varphi_1(+\infty) = \varphi_2(+\infty)$. In the retarded-advanced basis, these SK boundary conditions are equivalent to requiring that $\varphi_a(+\infty)=0$. As a result, there is just one copy of every global symmetry and every time-independent gauge-transformation. By contrast, there are two copies of gauge symmetries with arbitrary time dependence. 
\end{itemize}

\subsection{Non-equilibrium coset construction}


Suppose we have a $d+1$-dimensional relativistic system with global (zero-form) symmetry group $\cG$, which includes both internal and spacetime symmetries and is spontaneously broken to the subgroup $\cH$. We denote the generators by
\begin{equation}\begin{split}\label{generators}
\bar P_\mu &= \text{unbroken translations,}\\
T_A &= \text{other unbroken generators,}\\
\tau_\alpha & = \text{broken generators}.
\end{split}\end{equation}
The unbroken subgroup $\cH$ is then generated by $\bar P_\mu$ and $T_A$.
 We assume the existence of $d+1$ unbroken translation generators $\bar P_\mu$,  but do not require them to be the spacetime translation generators of the Poincaré group. 

We construct the effective action on the `fluid worldvolume' coordinates $\sigma^M$ for $M=0,\dots,d$ and parameterize the most general elements (one for each leg of the SK contour) of $\cG$ by
\begin{equation} g_s(\sigma) = e^{i X_s^\mu(\sigma)\bar P_\mu} e^{i\pi_s^\alpha(\sigma)\tau_\alpha} e^{ib^A_s(\sigma) T_A},~~~~~~~~~~s=1,2.  \end{equation}
Under a global symmetry transformation $\gamma\in \cG$, transformations of the Goldstone fields can be computed according to 
\begin{equation} g_s[X_s,\pi_s,b_s] \to g_s[X'_s,\pi'_s,b'_s] \equiv \gamma\cdot g_s[X_s,\pi_s,b_s]. \end{equation}

In order to distinguish the broken and unbroken symmetries, we require that the EFT be invariant under certain time-independent gauge transformations.\footnote{We say that these transformations are gauge symmetries because they do not change the state of the system, but are instead redundancies of description.} Throughout the rest of the paper, we let indices $M,N,P,Q=0,\dots,d$ be coordinate indices of $\sigma$ and we let $I,J,K,L=1,\dots,d$ be {\it spatial} coordinate indices of $\sigma$. First, since we are assuming the existence of $d+1$ translation generators,  impose the time-independent diffeomorphism symmetry on the coordinates
\begin{equation}\label{fluid sym}\sigma^0 \to \sigma^0 + f(\sigma^I),~~~~~~~~~~\sigma^I \to \Sigma^I(\sigma^J),\end{equation} 
for arbitrary $f$ and $\Sigma^I$. 
If there were fewer than $d+1$ unbroken translations, then this set of coordinate symmetries could be reduced; see~\cite{Landry second sound}. Next, because $T_A$ are unbroken, we have the right-acting, time-independent gauge symmetries
\begin{equation}\label{right-gauge sym} g_s(\sigma) \to g_s(\sigma)\cdot e^{i \lambda^A(\sigma^I) T_A}. \end{equation}

Finally, we compute the Maurer-Cartan form, which is a Lie algebra-valued one-form given by $g_s^{-1} d g_s$. Because it is Lie algebra-valued, it is always expressible as a linear combination of symmetry generators. In particular, we have
\begin{equation}\label{MC form}g_s^{-1} d g_s =  i d\sigma^M\squ{ E_{sM}^\mu \cur{ \bar P_\mu + i \nabla_{\mu} \pi_s^\alpha \tau_\alpha} +  \cB_{sM}^A T_A}. \end{equation}
It can be checked that under the gauge symmetries~(\ref{fluid sym}) and~(\ref{right-gauge sym}), $E_{sM}^\mu$ transform as vielbeins, $\nabla_{s\mu}\pi^\alpha$ transform linearly and hence we call them `covariant derivatives,' and $\cB_{sM}^A$ transform as gauge fields. The effective action is then constructed by forming manifestly symmetry-invariant terms out of the vielbeins, covariant derivatives, and connections. 

Finally, suppose that $\Theta$ is some anti-unitary, time-reversing symmetry of the ultraviolet theory and that the equilibrium density matrix describes a state in thermal equilibrium, namely
\begin{equation}\rho  = \frac{e^{-\beta_0 \bar P^0}}{\text{tr} e^{-\beta_0 \bar P^0}} , \end{equation}
where $\beta_0$ is the equilibrium inverse temperature. 
Then, the effective action must be invariant under the so-called dynamical KMS symmetries, which are non-local $\mathbb Z_2$ symmetries whose actions are\footnote{The factors $e^{\theta \bar P_0}$ and $e^{-(\beta_0-\theta) \bar P_0}$ arise because the field $X_s^0(\sigma)$ transform as a time coordinates under dynamical KMS symmetries. }
\begin{equation}\begin{split} \label{dynamical KMS} g_1(\sigma) &\to \Theta e^{\theta \bar P_0}g_1(\sigma^0 - i\theta , \vec \sigma),\\
g_2(\sigma) &\to \Theta e^{ -(\beta_0-\theta)\bar P_0 }g_2(\sigma^0 + i(\beta_0 -\theta) , \vec \sigma),  \end{split}\end{equation}
for arbitrary $\theta\in[0,\beta_0]$~\cite{H. Liu}. 
In the classical limit, these symmetries reduce to a single local $\mathbb Z_2$ transformation. 

To take the classical limit, it is convenient to work in the retarded-advanced basis.
In particular for a given set of fields $\varphi_s(\sigma)$, $s=1,2$ defined on the SK contour, we define the retarded-advanced fileds by
\begin{equation}\label{retarded-advanced}\varphi_r = \frac{1}{2}(\varphi_1+\varphi_2),~~~~~~~~~~\varphi_a = \varphi_1-\varphi_2. \end{equation}
It will be useful to define the $a$-derivative denoted by $\delta_a$ whose action is given by $\delta_a \varphi_r=\varphi_a$, for any retarded-advanced pair of fields $\varphi_r$ and $\varphi_a$. We also require that $\delta_a^2=0$ and that it satisfy the Leibnitz rule, namely
\begin{equation}\delta_a (\varphi_r \varphi_r') = \varphi_a \varphi_r' + \varphi_r \varphi_a' .\end{equation}

Then, the way to construct classical invariant building-blocks is as follows. Parameterize the most general group element with retarded Goldstones 
\begin{equation} g_r(\sigma) = e^{i X_r^\mu(\sigma)\bar P_\mu} e^{i\pi_r^\alpha(\sigma)\tau_\alpha} e^{ib^A_r(\sigma) T_A} , \end{equation}
and construct the Maurer-Cartan form $g_r^{-1} d g_r$, which is given by~\eqref{MC form} with replacement $s\to r$. Then, we can construct retarded building-blocks from this Maurer-Cartan form in the usual way. To construct advanced building-blocks, we need only act with $\delta_a$ on terms of the retarded Maurer-Cartan form. The classical dynamical KMS symmetry is then
\begin{equation}\label{classical KMS}
g_r(\sigma)\to \Theta g_r(\sigma),~~~~~g_a(\sigma) \to \Theta g_a(\sigma) - i \Theta \beta_0 \frac{\partial g_r(\sigma)}{\partial\sigma^0} + \Theta \beta_0 \bar P_0 g_r(\sigma)  ,
\end{equation} 
where $g_a\equiv \delta_a g_r$. 


\section{Non-Stückelberg symmetries}

In non-equilibrium systems, it is possible to have a symmetry with no corresponding non-trivial Noether current. As a simple example, consider a non-relativistic point particle in a uniform fluid undergoing Brownian motion. The effective action for the point particle is given by
\begin{equation}I_\text{p.p.} = \int dt \big[ -\nu x_a \dot x_r + \tilde M \dot x_r \dot x_a +\frac{i}{2}\sigma x_a^2 \big] ,\end{equation}
where the retarded field $x_{r}(t)$ gives the classical position of the particle, $x_a(t)$ is the corresponding advanced field, and the dynamical KMS symmetry requires that $\nu = \frac{1}{2} \sigma \beta_0 \geq 0$, which is just a statement of the fluctuation-dissipation theorem. 
Evidently, the physics of the point particle is independent of its position in space and time as the action is invariant under 
\begin{equation}t\to t+c_0,~~~~~~~~~~ x_r\to x_r+c_1,\end{equation}
for constants $c_0$ and $c_1$. 
In other words, it enjoys spacetime translation symmetry. Given that it exists in a medium, however, the particle is subject to friction and hence its energy and momentum are not conserved. In particular, the equations of motion are
\begin{equation} \tilde M \ddot x_r = -\nu \dot x_r.  \end{equation}
 Thus, there are no conserved currents associated with the spacetime translation symmetry at the level of the equations of motion.\footnote{Strictly speaking, there is a conserved current there, namely $\tilde M \dot x_r + \nu x_r$. The existence of such a conserved quantity is accidental, however, and it does to persist when higher-order terms are included.} This finding may puzzle some readers: we have an action with a global symmetry, meaning that the assumptions of Noether's theorem are satisfied. So how can there be no conserved current? The answer is that at the level of the mathematics, there is a conserved current associated with every symmetry, but it need not have physically meaningful dynamics. To see how this is so, let us compute the Noether current associated with spatial translations. We have\footnote{Notice that $j$ has no indices because we are working with a $0+1$-dimensional QFT. }
\begin{equation} j = -\nu x_a + \tilde M \dot x_a.  \end{equation}  
But when the equations of motion are satisfied, all advanced terms vanish, so we have $x_a=0$ and as a result $j=0$. Thus the conservation equation $\partial_t j=0$, while true, provides no physical information about the system. It should be noted, however, that wile this current is trivial at the level of equations of motion, it will lead to Ward identities when correlation functions are considered.

At the level of non-equilibrium EFTs, the existence of a conserved current is closely related to gaugeability. In particular, to ensure that a particular current is conserved in a non-equilibrium EFT, we must introduce external gauge fields as well as dynamical Stückelberg fields. In the present case, the fields $x_r$ and $x_a$ are not Stückleberg fields for the metric (or any other external gauge source). As a result, their equations of motion do not yield conservation laws. For further explanation, see \S\ref{Gauge fields and the Maurer-Cartan form} or consult~\cite{Landry quasicrystal,Landry second sound, Landry chemistry,Landry}. As mentioned in the introduction, we refer to symmetries with corresponding current conservation as {Stückelberg symmetries} and those without conserved current as {non-Stückelberg symmetries} (NSS).

In the point-particle example, the fact that space and time translations are NSS owes to the fact that the particle is an open system and can freely exchange energy and momentum with an environment. Such behavior, however, does not always arise because of interactions with an environment. For example, the diffusive phason mode in a quasicrystal owes its existence to a $U(1)$ NSS with no corresponding conserved current~\cite{Landry quasicrystal}. We will see other examples of NSS that arise in closed systems in subsequent sections of this work.

It can be shown that if we build the the effective action exclusively from terms furnished by the coset construction, then our Goldstone fields are gaugeable; see \S\ref{Gauge fields and the Maurer-Cartan form}. As a result, the fields furnished by the cosets are secretly Stückelberg fields in which the external gauge fields have been set to trivial background values. This observation raises the question: what kinds of terms are invariant under the global symmetry group but are not gaugeable? That is, how do we used a coset construction method to formulate actions with NSS? To answer this question, consider 
\begin{equation}g_a(\sigma) \equiv g_2^{-1}(\sigma)\cdot g_1(\sigma). \end{equation}  
Evidently  $g_a$ transforms by conjugation under the gauge symmetry defined by~(\ref{right-gauge sym}) and is invariant under all global (i.e. physical) $\cG$-symmetry transformations. Notice that $g_a\in\cG$, while the Maurer-Cartan forms are Lie-algebra valued. To turn $g_a$ into a Lie algebra-valued object, we need only take the log, that is
\begin{equation} \label{advanced form 0} \bG_a \equiv \log g_a.  \end{equation}
We can then expand $\bG_a$ as a linear combination of symmetry generators 
\begin{equation}\bG_a =i \Phi_{\bar P}^\mu \bar P_\mu + i\Phi_\tau^\alpha \tau_\alpha + i \Phi_T^A T_A.  \end{equation}
If a given symmetry is a NSS, we must include the corresponding $\Phi$ as a building-block in the effective action. For example if we want $\tau_\alpha$ to be a NSS, then we must use $\Phi_\tau^\alpha$ as a covariant building-block in the EFT. Conversely if we want $\tau_\alpha$ to be a Stückelberg symmetry, then we may not use $\Phi_\tau^\alpha$ as a building-block. 

Taking the classical limit, we have $\bG_a = g_r^{-1} \delta_a g_r$. Thus, $\bG_a$ becomes like a Maurer-Cartan form with respect to the ``derivative'' defined by $\delta_a$. 

\subsection{Reactive fluids}

We will now demonstrate how to use the coset construction to formulate an EFT for fluids that exhibit reactive flows. 
There are, in addition to Poincaré symmetry, $n$ different unbroken $U(1)$ charges that we denote by $N_1,\dots,N_n$. Each $U(1)$ charge counts the number of particles of a given species. To allow for chemical reactions, that is, the transformations of certain types of particles into other types of particles, we must kill the conservation of some subset of these charges. But first, we will construct the action for which there are no chemical reactions.

It turns out that we will need to use the full SK contour to describe these dynamics, so we must construct an action with doubled field content. We will work in the classical limit, so we can use the $\delta_a$ trick to simplify matters. The most general retarded group element is 
\begin{equation}g_r(\sigma) = e^{i X_r^\mu(\sigma) P_\mu} e^{i \psi^A_r(\sigma) N_A} e^{i \eta_r^i(\sigma) K_i} e^{i\theta_r^i(\sigma) J_i}, \end{equation} 
where $A=1,\dots,N$. 
Because translations are unbroken, we impose the fluid diffeomorphism symmetry~(\ref{fluid sym}) and because $J_i$ and $N^A$ are unbroken we have the gauge symmetries
\begin{equation} g_r(\sigma)\to g_r(\sigma) \cdot e^{i \lambda^i(\sigma^I) J_i},~~~~~~~~~~ g_r(\sigma)\to g_r(\sigma) \cdot e^{i c^A(\sigma^I) N_A}. \end{equation}

The resulting retarded Maurer-Cartan form is then
\begin{equation}g_r^{-1} d g_r = i  E_r^\mu (P_\mu + \nabla_{\mu} \eta_r^i K_i) +i \omega_{r}^i J_i + i \cB_{r}^A N_A, \end{equation} 
where
\begin{equation}\begin{split}
E_{rM}^\mu & =  \partial_M X_r^\nu [\Lambda_r R_r{]_\nu}^\mu ,\\
\nabla_\mu \eta_r^i & = (E^{-1}_r)_\mu^M[\Lambda_r^{-1}\partial_M \Lambda_r ]^{0j} R_r^{ji}, \\
\omega_{rM}^i & = \frac{1}{2}  \epsilon^{ijk} [(\Lambda_r R_r)^{-1} \partial_M (\Lambda_r R_r)]^{jk}, \\
\cB_{rM}^a & =  \partial_{M}\psi_r ,
\end{split}\end{equation}
such that ${\Lambda_r^\mu}_\nu = (e^{i\eta_r^i K_i}{)^\mu}_\nu$ and $R_r^{ij} = (e^{i\theta_r^i J_i})^{ij}$. 

To remove Lorentz Goldstones, we impose the IH constraints 
\begin{equation} E_{r0}^i = 0 \implies\frac{\partial_0 X^i}{\partial_0 X^t}= - \frac{\eta_r^i}{\eta_r}\tanh \eta_r,\end{equation}
and 
\begin{equation} \epsilon^{ijk} (E^{-1}_r)^M_j E_{aM}^k = 0,~~~~~ \epsilon^{ijk} (E^{-1}_r)^M_j \partial_0 E_{rM}^k = 0,\end{equation}
where  $E_{aM}^\mu=\delta_a E_{rM}^\mu$. This second IH constraint can be solved to remove the rotation Goldstones, but it is not necessary at this level in the derivative expansion; see~\cite{Landry}. 

After imposing these IH constraints, we are left with the invariant building-blocks
\begin{equation} 
T = \frac{1}{\sqrt{-G_{r00}}},~~~~~~~~~~\mu^A = \frac{1}{\sqrt{- G_{r00} } } \frac{\partial \psi_r^A}{\partial\sigma^0},
\end{equation} 
and the invariant integration measure is
\begin{equation}d^4 \sigma \sqrt{- G_r} . \end{equation}
As before, we have defined the fluid metric by $G_{rMN} = E_{rM}^\mu \eta_{\mu\nu} E_{rN}^\nu$. 

To construct advanced building-blocks, we need only act with $\delta_a$ on terms from the retarded Maurer-Cartan form. At leading order, the dynamical KMS symmetries allow us to write
\begin{equation}I_\text{EFT} = \delta_a S, \end{equation}
where 
\begin{equation} S = \int d^4 \sigma \sqrt{- G_r} P(T,\mu^A). \end{equation}
At this point, it is convenient to transform to the physical spacetime, so letting $x^\mu\equiv X_r^\mu$, we have\footnote{The fluid coordinates are now dynamical fields $\sigma^M(x)$.}
\begin{equation} S = \int d^4 x \, P(T,\mu^A). \end{equation}
To be explicit about the form of the non-equilibrium effective action, we may write 
\begin{equation}\label{no reactions} I = \int d^4 x \squ{ T^{\mu\nu} \partial_\mu X_{a\nu} + J^{A\mu} \partial_\mu \psi^A_a} ,\end{equation}
where the stress-energy tensor is
\begin{equation}T^{\mu\nu} = \epsilon(T,\mu^A) u^\mu u^\nu + p(T,\mu^A) \Delta^{\mu\nu},\end{equation}
such that $\Delta^{\mu\nu} = \eta^{\mu\nu}+u^\mu u^\nu$ and the $U(1)$ currents associated with $N^A$ are
\begin{equation}J^{A\mu} = n^A(T,\mu^B) u^\mu . \end{equation}
The (classical) dynamical KMS symmetry imposes the relations
\begin{equation}  \begin{split}
\epsilon + p & = T\frac{\partial p}{\partial T}+\mu^A \frac{\partial p}{\partial \mu^A}, \\
n^A & = \frac{\partial p}{\partial \mu^A} ,
\end{split} \end{equation}
which are equivalent to the local first law of thermodynamics. 
The equations of motion are just the conservation of energy, momentum and charge, namely
\begin{equation}\partial_\mu T^{\mu\nu} = 0 ,~~~~~~~~~~ \partial_\mu J^{A\mu} = 0. \end{equation}
Since $J^{A\mu}$ are all independently conserved, we see that our action describes fluids with $n$ species of particles that are not permitted to undergo any chemical reactions.

To include chemical reactions, we must kill certain linear combinations of $U(1)$ conservation equations, while preserving other linear combinations. Such non-conservation can be accomplished by treating certain linear combinations of the $U(1)$ symmetries as as NSS. Without loss of generality, let $N^{\hat A}$ for $\hat A=1,\dots, k<n$ be representatives of the coset of non-conserved $U(1)$ charges. Then we have invariant building-blocks furnished by
\begin{equation} g_r^{-1} \delta_a g_r \supset i \psi_{\hat A} N_{\hat A}. \end{equation} 
We therefore find that new terms can be added to~\eqref{no reactions} and obtain
\begin{equation}\begin{split} \label{with reactions} I = \int d^4 x \bigg[ T^{\mu\nu} \partial_\mu X_{a\nu} + J^{A\mu} \partial_\mu \psi^A  - \Gamma^{\hat A} \psi_a^{\hat A} + \frac{i}{2} M^{\hat A\hat B}\psi^{\hat A}_a \psi^{\hat B}_a  \bigg] ,\end{split}\end{equation}
where the dynamical KMS symmetry imposes
\begin{equation} \Gamma^{\hat A} = \frac{1}{2} M^{\hat A \hat B} \beta^\mu \partial_\mu \psi_r^{\hat B} . \end{equation}
The equations of motion for $X_a^\mu$ are still the conservation of the stress-energy tensor. But the equations for $\psi_a^A$ are now
\begin{equation}\partial_\mu J^{A\mu} = - \Gamma^A ,\end{equation}
where $\Gamma^A\equiv \delta^{\hat A A}  \Gamma^{\hat A} $. But this is just the equation for $k$ non-conserved $U(1)$ currents and $n-k$ conserved $U(1)$ currents. Thus, chemical reactions may take place. These results agree with those of~\cite{Landry chemistry}.

\section{Higher-form symmetries and the coset construction}

When a zero-form symmetry is spontaneously broken, there exists a continuum of vacuum states. The value of the Goldstone field at a given point $A_0$ can be interpreted as defining the local vacuum near that point. If we move to a different point, $B_0$, then there will be a new local vacuum that is related to the local vacuum at $A_0$ by a symmetry transformation. The Maurer-Cartan form tells us how the local vacuum changes from point to point. In particular, suppose $\Omega_1$ is the Maurer-Cartan one-form and $V_1$ is a path connecting $A_0$ and $B_0$. In other words, $\partial V_1 = A_0\cup (-B_0)$.\footnote{The minus sign on $B_0$ indicates a reversal of orientation. } We can represent the local vacuum at arbitrary point $x$ by an element of the symmetry group $g_0(x) \in \cG$. Then, we have
\begin{equation} \cP e^{ \int_{V_1} \Omega_1 } = g_0^{-1}(B_0) g_0(A_0), \end{equation}
where $\cP$ indicates path-ordering. 

Higher-form symmetries admit a straight-forward generalization of the Maurer-Cartan form. In particular, given a spontaneously broken $p$-form symmetry, let $A_p$ and $B_p$ be $p$-dimensional surfaces and $V_{p+1}$ be a $p+1$-dimensional surfaces such that $\partial V_{p+1} = A_p \cup (-B_p)$. Then, we define the Maurer-Cartan $p+1$-form $\Omega_{p+1}$ by 
\begin{equation} e^{ \int_{V_{p+1}} \Omega_{p+1} } = g_p^{-1}(B_p) g_p(A_p), \end{equation}
where $g_p^{-1}(A_p)= e^{i \bQ_p \int _{A_p} \varphi^p}$ and $g_p^{-1}(B_p)= e^{-i \bQ_p \int _{B_p} \varphi^p}$, such that path-ordering is implicit for $p=1$, are $p$-dimensional generalizations of Wilson lines. Here, $\varphi^p$ is the $p$-form Goldstone corresponding to the $p$-form symmetry charge $\bQ_p$. 
Notice that for $p>0$ we have no notion of path-ordering for the l.h.s., which is related to the fact that all $p>0$-form symmetries are abelian; however, if $p=1$ then the r.h.s. involves Wilson lines, which must be path-ordered. In either case, we have for $p>0$, 
\begin{equation} e^{ \int_{V_{p+1}} \Omega_{p+1} } = g_p(\partial V_{p+1}),  \end{equation}
where the r.h.s. is path-ordered if $p=1$.

\begin{figure}
\centering
\begin{tikzpicture}
\draw[fill=blue,draw opacity=0] (-0.02,0) to[bend left=80]node[anchor=south] {$B_p$} (6.02,0) to (5.8,0) to[bend right=79.8] (.2,0) to (-.02,0) ;
\draw[fill=red,draw opacity=0] (-0.02,0) to[bend right=80]node[anchor=north] {$A_p$} (6.02,0) to (5.8,0) to[bend left=79.8] (.2,0) to (-.02,0) ;
\draw[fill=gray!15,draw opacity=0]    (.1,0)  to[bend left=78] (5.9,0) to[bend left=78]  (.1,0);
\draw[black] (3,-.3) -- node[anchor=south] {$V_{p+1}$} (3,-.3);
\end{tikzpicture}
\caption{ This figure depicts a $p+1$-dimensional surface $V_{p+1}$ (gray) and two $p$-dimensional surfaces $A_p$ (red) and $B_p$ (blue). The boundary of $V_{p+1}$ is equal to the union of $A_p$ and $B_p$, that is $\partial V_{p+1} = A_p\cup (-B_p)$.  }
\end{figure}
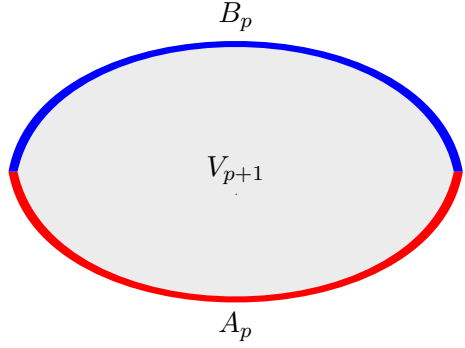

Supposing we have a mixture of various higher-form symmetries of differing $p$. Let $A = \{A_p\}_{p=0}^d$, $B = \{B_p\}_{p=0}^d$, and $V= \{V_{p+1}\}_{p=0}^d$ such that $\partial V_p = A_p\cup (-B_p)$ for all $p$. 
Then, we define the Maurer-Cartan to be the mixed-form $\Omega = \sum_p \Omega_p$ such that 
\begin{equation} e^{ \int_{V} \Omega } = g^{-1}(B) g(A), \end{equation}
where we have left the path-ordering for one-dimensional integrals implicit and we define 
\begin{equation} \int_{V} \Omega\equiv \sum_p \int_{V_p} \Omega_p,~~~~~~~~~~g(A) \equiv \prod_p g_p(A_p),~~~~~~~~~~g(B) \equiv \prod_p g_p(B_p) .\end{equation}

We are primarily interested in non-equilibrium systems, so we must double field content and define our action on the fluid coordinates $\sigma$. As usual we let $s=1,2$ indicate on which leg of the SK contours the fields live. 
Thus letting the zero-form symmetry generators be given by~\eqref{generators} and letting $\bQ_p$ be the $p>0$-form charges, we have\footnote{We choose $\bQ_p$ to be defined on a surface that ``links'' with $\Sigma_p$.  }
\begin{equation}\label{group elt form}g_s(\Sigma) = \cur{\prod_{p>0} e^{i\bQ_p \int_{\Sigma_p} \varphi^p_s } } e^{i X_s^\mu(\sigma) \bar P_\mu} e^{i\pi_s^\alpha(\sigma)\tau_\alpha} e^{ib^A_s(\sigma) T_A} ,  \end{equation}
where we used $\Sigma = \{\Sigma_p\}_{p=0}^d$ and we identified $\sigma= \Sigma_0$. 
If a given $\bQ_p$ is not spontaneously broken then we impose the time-independent gauge symmetry
\begin{equation} \label{higher form gauge 1} g_s(\Sigma) \to g_s(\Sigma) \cdot e^{i \bQ_p \int _{\Sigma_p}\kappa^p },  \end{equation} 
where $\kappa^p= \kappa^p(\sigma^I)$ for $I=1,\dots,d$ is a time-independent $p$-form such that $\kappa^p_{0 M_2\dots M_p} = 0$.

With the group elements~\eqref{group elt form} in mind, we may now compute the Maurer-Cartan mixed-forms  
\begin{equation}\begin{split} \label{MC form p} \Omega_s =  i  E_{s}^\mu \cur{ \bar P_\mu + i \nabla_{s\mu} \pi^\alpha \tau_\alpha} +  \cB_{s}^A T_A  + \bF^p_{s} \bQ_p , \end{split}\end{equation}
where, $E_s^\mu =E_{sM}^\mu d\sigma^M$, $\cB_s^A = \cB_{sM}^A d\sigma^M$ and $\bF^p_s = \frac{1}{(p+1)!}\bF^p_{s M_0\dots M_p} d\sigma^{M_0}\wedge \cdots\wedge d\sigma^{M_p}$. The $p>0$-form symmetry associated with charge $\bQ_p$ acts by
\begin{equation}\label{higher form gauge} g_s(\Sigma) \to e^{i \bQ_p \int_{\Sigma_p }\alpha_s^p(\sigma)} \cdot  g_s(\Sigma) , \end{equation}
for any closed $p$-forms $\alpha_s^p$. Notice that since $\alpha_s^p$ can have arbitrary time-dependence, the SK future-time boundary conditions permit the $s=1,2$ transformations to be different; however, all topological properties of $\alpha_1^p$ and $\alpha_2^p$ must be the same. Thus, $\alpha_1^p-\alpha_2^p$ must be exact. In this way, there is just one copy of each global $p$-form symmetry.   
Finally, once the Maurer-Cartan form is computed, we can go through the usual procedure of constructing manifestly-invariant terms built from the coefficients of the Maurer-Cartan form.

\subsection{The zero-temperature limit}

There are two meanings of the zero-temperature limit. The first is the most straight-forward. We merely formulate an effective action using the in-out formalism; as a result we only have one set of fields (as opposed to the SK doubled field content). The second meaning, by contrast, involves a thermal system at very low temperature such that all temperature fluctuations can be neglected. Essentially, we are interested in a regime in which thermal activity is suppressed, but not truly absent. As a result, we still need the SK contour and doubled field content. To distinguish these two possibilities, we use $T=0$ to indicate the true zero-temperature case and $T\to 0$ to indicate the very small temperature case. \\

\noindent {\bf T=0:} The coset construction for systems at zero temperature is well-studied, so the only novel comments we can make have to do with higher-form symmetries. 
With the zero-form generators given by~(\ref{generators}) and the $p>0$-form generators $\bQ_p$, the most general coset element~is 
\begin{equation} g(\Sigma) =\cur{\prod_{p>0} e^{i\bQ^p \int_{\Sigma_p} \varphi^p_s } }e^{i x^\mu \bar P_\mu} e^{i\pi^\alpha(x) \tau_\alpha} ,  \end{equation} 
where we identify $x=\Sigma_0$ and we have supposed that $\bQ_p$ are spontaneously broken. 
Then, the Maurer-Cartan mixed-form is defined in the usual way and we can extract from it symmetry-covariant building-blocks. Notice that in the zero-temperature limit, there is no possibility of having unbroken Goldstones or symmetries without corresponding Noether currents.
 \\

\noindent {\bf T$\to$0:} Now we take the small-temperature limit. To do this, we retain all of the same machinery as the finite-temperature case, but with two small alterations. We begin by modifying the fluid symmetry~(\ref{fluid sym}). To determine how to modify this diffeomorphism symmetry, we must first understand where it comes from.  Suppose that we have full diffeomorphism symmetry $\sigma^M \to \Sigma^M(\sigma)$. For unbroken translations, this is the most natural symmetry to suppose because it allows us to gauge-fix by $\sigma^M = x^\mu \delta^M_\mu$, which means our EFT is defined on the physical spacetime in the usual way~\cite{Landry spin}.  At finite temperature, we introduce the inverse temperature four-vector field $\beta^M(\sigma)$, which transforms as a contravariant vector under diffeomorphisms of the fluid worldvolume. Under a coordinate transformation, we have
\begin{equation} \beta^M \to \beta^N \frac{\partial \Sigma^M}{\partial \sigma^N} .\end{equation}
Thus, if we gauge-fix $\beta^M = \beta_0 \delta_0^M$, we find that the residual gauge symmetries are~\eqref{fluid sym}, as expected. But now we wish to take the zero-temperature limit, which is equivalent to taking $\beta_0\to \infty$. As a result, the magnitude of $\beta^M$ is no longer meaningful, but the direction is still important. Therefore we impose the gauge-fixing condition $\beta^M \propto \delta^M_0$, where we allow transformations that alter the magnitude of $\beta^M$ without changing its direction. We are thus left with the residual gauge symmetries
\begin{equation} \sigma^0 \to \sigma^0 + f(\sigma^0,\sigma^I),~~~~~~~~~~\sigma^I \to \Sigma^I(\sigma^J).  \end{equation} 
We take this expanded set of symmetries to be the low-temperature fluid worldvolume symmetries. 

Now we must modify the dynamical KMS symmetries. In the $\beta_0\to 0$ limit, the transformations~\eqref{dynamical KMS} are no longer well-defined. Instead, we impose continuous symmetries of the form~\eqref{dynamical KMS} or in the classical limit~\eqref{classical KMS} but now allow $\beta_0$ to be a continuous real parameter. Notice that if we take $\beta_0=\theta=0$, we find that our EFT is invariant under $\Theta$ alone. Therefore our action is invariant under the following symmetries that do not involve time reversal
\begin{equation}\begin{split} \label{T to 0 dynamical KMS} g_1(\sigma) &\to  g_1(\sigma^0 - i\theta , \vec \sigma),\\
g_2(\sigma) &\to  g_2(\sigma^0 + i(\beta_0 -\theta) , \vec \sigma),  \end{split}\end{equation}
for arbitrary $\theta$ and $\beta_0$ such that $|\theta| \leq \beta_0$. In the classical limit, we have 
\begin{equation} 
g_r \to g_r,~~~~~g_a \to  g_a + i  \beta_0 \frac{\partial g_r}{\partial\sigma^0} ,
\end{equation}
for arbitrary $\beta_0$. Notice that this classical KMS symmetry is a $U(1)$ symmetry that acts only on the advanced fields. As a result, there must be a retarded Noether current that is conserved.\footnote{The fact that the Noether current is retarded means that it is physically important at the level of the equations of motion.} This Noether current has the interpretation of entropy and is conserved in the $T\to 0$ limit, as expected. In the $T\neq 0$ limit, it is no longer conserved as the dynamical KMS symmetries are no longer continuous~\cite{H. Liu}.

\subsection{Gauge theories}

In this subsection, we will construct the actions for systems involving gauge bosons. We begin by woking at zero-temperature and construct the familiar pure Yang-Mills $SU(N)$ gauge theory. Then we construct the Chern-Simons action, a purely topological theory, from symmetry considerations alone. We conclude by turning attention to finite temperature systems and reproduce the effective action for magnetohydrodynamics that was first presented in~\cite{MHD,Armas:2018ibg,Armas:2018atq,Armas:2018zbe}. Strictly speaking, we should only use the higher-form coset construction to formulate abelian gauge theories as only these gauge theories describe continuous higher-form symmetries. But it is nevertheless an interesting exercise to push the coset construction outside of its expected regime of validity and discover that we arrive at the correct answers anyway. 

\subsubsection{Yang-Mills}

As a simple example, we construct the action for pure Yang-Mills in $3+1$ dimensions using this new, higher-form coset construction. We consider the zero-temperature, $T=0$, case, meaning there is only one copy of the fields. To construct an action with a higher-form symmetry, we must define the Wilson loops. Let $t_a$ be the generators of a compact Lie group $G$ such that
\begin{equation}\label{G generators} [t_a,t_b] = i {f_{ab}}^c t_c, \end{equation}
where ${f_{ab}}^c$ are the structure constants. We define the Wilson loop by
\begin{equation} g(C) = \cP e^{i q \int_C A} ,\end{equation}
where $C$ is a closed path, $q$ is the charge of the Wilson loop, and $A \equiv A^a_\mu dx^\mu t^a$. For Yang-Mills theory, we take $G=SU(N)$; then we will have a global $\mathbb Z_n$ one-form symmetry~\cite{Harlow:2018tng}.\footnote{Ordinarily, the coset construction cannot be used when the symmetry group is discrete as there are no Goldstone modes. There is no guarantee that naïvely applying the coset construction to this case should work, but we will nevertheless proceed and find that everything works out. We should note, however, that unlike the abelian case, the gauge field $A$ does not have the interpretation as a Goldstone as no continuous symmetry is spontaneously broken. These remarks apply equally to the next example in which we consider Chern Simons theory.} Letting $S$ be a two-dimensional surface with $\partial S=C$, define the Maurer-Cartan two-form $\Omega_2$ implicitly by
\begin{equation}e^{i \int_S \Omega_2} = \cP e^{i q \int_{\partial S} A}. \end{equation}
Taking $S$ infinitesimal, we can expand both sides of the above equation to find 
\begin{equation}\begin{split} \int_{S} \Omega_2 = i q \int_{\partial S} A - \frac{q^2}{2} \int_0^1 d\lambda_l \int_0^1 d\lambda_2 \frac{d z^\mu}{d\lambda_1} \frac{d z^\nu}{d\lambda_2} A^a_\mu A^b_\nu \cur{t^a t^b \theta(\lambda_1-\lambda_2) + t^b t^a \theta(\lambda_2-\lambda_1)}. \end{split}\end{equation}
Using Stoke's theorem, the r.h.s. can be simplified to yield  
\begin{equation} \int_{S} \Omega_2  = i q  \int _{S} \cur{dA + i q A\wedge A }, \end{equation} 
indicating that $\Omega_2 = i q (dA+i q A\wedge A)$. Recognize $F\equiv dA+iqA\wedge A$ as the field strength tensor for Yang-Mills theory. At leading order in the derivative expansion, there is only one term that is invariant under the global higher-form symmetry, namely 
\begin{equation} S = -\frac{1}{4} \int d^4 x \,F^a_{\mu\nu} F^{a\mu\nu}, \end{equation}
where the factor of $-1/4$ is a convention that fixes field normalization. We have thus constructed the familiar Yang-Mills action.

\subsubsection{Chern Simons}

One of the original motivations for considering higher-form symmetries was to understand topological phases of matter from symmetry principles alone. Here we will construct the leading-order action for Chern-Simons theory in $2+1$ dimensions. We will consider a zero-temperature system so we need only one copy of the fields. 

In addition to Poincaré symmetry, which is unbroken, we postulate the existence of a spontaneously broken $U(1)$ one-form symmetry with Goldstone $A=A_\mu dx^\mu$. Inspired by the success of the previous subsection, we promote the $U(1)$ gauge symmetry of $A$ to a more general, non-abelian gauge symmetry. Let~\eqref{G generators} be the generators of the compact Lie group $G$ such that $A=A_\mu^a dx^\mu t^a$.  The Maurer-Cartan two-form, $\Omega_2$, is then defined by\footnote{We drop the factor of $i$ and the charge from the exponent on the r.h.s. so that our conventions match standard results. }
\begin{equation} e^{ \int_{\Sigma_2} \Omega_2} = e^{\int_{\partial\Sigma_2} A}.  \end{equation}
Following the Yang-Mills construction, we arrive at the covariant building-block $F=dA+A\wedge A$, which we interpret as the field strength. 

Chern-Simons theory involves a term that is  symmetry-invariant up to total derivative terms. Unfortunately, the coset construction only furnishes terms that are exactly symmetry-invariant. To circumvent this difficulty, we can construct the necessary term by working in $3+1$ dimensions and only consider terms that are total derivatives. The leading-order such $3+1$-dimensional action is then
\begin{equation} S_{3+1} = \frac{k}{4\pi}\int _{\cM_{3+1}} \text{tr}(F\wedge F) = \frac{k}{4\pi} \int_{\cM_{3+1}}\text{tr} \squ{d \bigg(dA+\frac{2}{3} A\wedge A\wedge A \bigg) },  \end{equation}
where $\cM_{3+1}$ is a $3+1$-dimensional manifold and $k$ is a phenomenological constant. 
Then, using Stoke's theorem, we obtain the $2+1$-dimensional action 
\begin{equation}S_{2+1} = \frac{k}{4\pi} \int _{M_{2+1}} \text{tr} \cur{A\wedge dA +  \frac{2}{3} A\wedge A\wedge A}, \end{equation}
for $2+1$-dimensional manifold $\cM_{2+1}$. 

\subsubsection{Magnetohydrodynamics}

Magnetohydrodynamics (MHD) is the study of electromagnetism and fluids. In particular, it is the study of fluids that can support the flow of electric current but cannot support electric charge density. Any regions of non-vanishing charge density locally equilibrate to zero exponentially fast. We will restrict our considerations to MHD in $3+1$ dimensions. Let $A_\mu$ be the photon field, $F_{\mu\nu}$ be the field strength given by $F=dA$, and $j^\mu$ be the electromagnetic four-current. Then the electromagnetic equations of motion are 
\begin{equation} \partial_\nu F^{\mu\nu} = j^\mu.  \end{equation}
One might be tempted to say that the symmetries of electromagnetism are the gauged $U(1)$ symmetry, but gauge symmetries are mere redundancies of description and are therefore unphysical. Because this zero-form $U(1)$ symmetry is not a genuine symmetry of nature, it need not survive in our EFT. The true symmetry of electromagnetism is the one-form $U(1)$ symmetry with corresponding current $\bJ = \star F$, or in index notation, 
\begin{equation}\bJ^{\mu\nu} = \frac{1}{2} \epsilon^{\mu\nu\lambda\rho}F_{\lambda\rho}.  \end{equation}
Notice that $\partial_\mu \bJ^{\mu\nu}=0$ is a mathematical identity meaning this current is automatically conserved off-shell. Thus, at the level of the coset construction, we have a one-form symmetry generated by some $\bQ$. The SSB pattern is simply that boosts are spontaneously broken and all other symmetries including $\bQ$ are unbroken. 

We are only interested in constructing the leading-order action, which, it turns out, factorizes as the difference of two ordinary actions. As a result, we will use just one copy of the fields. The most general group element is
\begin{equation}g(\Sigma) = e^{i X^\mu(\sigma) P_\mu} e^{i \eta^i(\sigma) K_i} e^{i \theta^i(\sigma) J_i} e^{i \bQ \int_{\Sigma_2}\varphi } . \end{equation}
Because translations are unbroken, we have the fluid symmetries~(\ref{fluid sym}) and because rotations and the one-form symmetry are unbroken, we have the time-independent symmetries
\begin{equation} g(\Sigma)\to (\Sigma) \cdot e^{i \lambda^i(\sigma^I) J_i},~~~~~~~~~~g(\Sigma)\to g(\Sigma) \cdot e^{i \bQ\int _{\Sigma_2}\kappa }, \end{equation}
where $\lambda^i(\sigma^I)$ is arbitrary and $\kappa=\kappa_M(\sigma^I) d\sigma^M$ is a time-independent one-form such that $\kappa_0=0$. 
And finally, we have the one-form symmetry $\varphi\to \varphi +\alpha(\sigma)$, for any closed one-form~$\alpha$. 

From this group element, we find that the Maurer-Cartan form is 
\begin{equation}\Omega  = i  E^\mu (P_\mu + \nabla_\mu \eta^i K_i) +i\Omega^i J_i + i\bF \bQ , \end{equation} 
where
\begin{equation}\begin{split}
E_M^\mu & = \partial_M X^\nu [\Lambda R{]_\nu}^\mu ,\\
\nabla_\mu \eta^i & = (E^{-1})_\mu^M[\Lambda^{-1}\partial_M \Lambda ]^{0j} R^{ji}, \\
\Omega_M^i & = \frac{1}{2}  \epsilon^{ijk} [(\Lambda R)^{-1} \partial_M (\Lambda R)]^{jk}, \\
\bF_{MN} & = \partial_{[M}\varphi_{N]},
\end{split}\end{equation}
such that ${\Lambda^\mu}_\nu = (e^{i\eta^i K_i}{)^\mu}_\nu$ and $R^{ij} = (e^{i\theta^i J_i})^{ij}$. 


We can now impose IH constraints to remove the Lorentz Goldstones. First, we remove the boost Goldstones by setting 
\begin{equation}\label{boost IH} E_0^i=0,\end{equation}
which can be solved to give
\begin{equation} \frac{\eta^i}{\eta}\tanh \eta = -\frac{\partial_0 X^i}{\partial_0 X^t}, \end{equation}
where $\eta \equiv \sqrt{\vec\eta^2}$. Second, to remove the rotation Goldstones, we may fix 
\begin{equation}\label{rotation IH}\epsilon^{ijk}(E^{-1})^M_i \partial_0 E_M^j = 0.\end{equation}
This IH constraint can be used to remove $\theta^i$ from the invariant building-blocks, but the solution is not necessary for the construction of the leading-order action. For the sake of brevity, we therefore will not solve it; interested readers can consult~\cite{Landry} for more details.

With these IH constraints imposed, the leading-order building-blocks are as follows. 
Define the fluid worldvolume metric by 
\begin{equation}\label{WV metric}G_{MN} = E_M^\mu\eta_{\mu\nu}E^\nu_N = \frac{\partial X^\mu}{\partial \sigma^M}\eta_{\mu\nu} \frac{\partial X^\nu}{\partial \sigma^N}.  \end{equation}
Then, we have that 
\begin{equation}\label{local temperature}T=\frac{1}{E_0^t} = \frac{1}{\sqrt{-G_{00}}}\end{equation}
is an invariant building-block, which can be interpreted as the local temperature. 
Additionally, we have the invariant building-block  
\begin{equation} \mu =\sqrt{ \bF_{0M} G^{MN} \bF_{0N}}, \end{equation}
where $G^{MN}$ is the inverse of the fluid worldvolume metric. We may interpret $\mu$ as the local chemical potential. There are no other invariant building-blocks at this order in the derivative expansion. Performing a coordinate transformation from $\sigma^M$ to the physical spacetime $x^\mu \equiv X^\mu$, we have the leading-order action 
\begin{equation} S = \int d^4 x \, P(T,\mu), \end{equation} 
which exactly matches the results of~\cite{MHD}. 

Finally, we take the $T\to 0$ limit, which can be accomplished by expanding the fluid worldvolume symmetry and allowing full time diffeomorphisms, namely $\sigma^0\to \sigma^0 +f(\sigma^0,\sigma^I)$. The only effect this enlarged gauge symmetry has is the removal of $T$ as an invariant building-block. We therefore have
\begin{equation} S_{T\to 0} = \int d^4 x \, P(\mu),\end{equation} 
which again matches the results of~\cite{MHD}.

\section{Higher-form (non)-Stückelberg symmetries and SSB}\label{Higher-form symmetries and t Hooft anomalies}

\subsection{Gauge fields and the Maurer-Cartan form}\label{Gauge fields and the Maurer-Cartan form}

We previously stated that the way to ensure all symmetries have corresponding conserved Noether currents is to construct the action using two distinct Maurer-Cartan one-forms---one for each leg of the SK contour---that transform under a single copy of the global symmetry group $\cG$. In this section, we will use a Stückelberg trick inspired by~\cite{H. Liu, Landry} to demonstrate that this approach is correct. 

We begin by introducing sources for the Noether currents corresponding to each symmetry generator of $\cG$, which amounts to introducing external gauge fields. Let the zero-form symmetry generators be
\begin{equation}\begin{split} \bar P_m & =\text{unbroken translations},
\\ T_A & = \text{other unbroken generators},
\\ \tau_\alpha & =\text{broken generators},
 \end{split}\end{equation}
 and let 
 \begin{equation}\bQ_p = p>0\text{-form charges.~~~~~~~~~~} \end{equation}
We now use $m,n=0,\dots,d$ to denote Lorentz indices and $\mu,\nu=0,\dots,d$ to denote physical spacetime coordinate indices.\footnote{It is now necessary to distinguish between Lorentz indices $m,n$ and physical spacetime coordinate indices $\m,\n$ because the Stückelberg trick requires that we gauge all symmetries including Lorentz. } We wish to introduce external sources corresponding to each of these symmetries so that we can construct a generating functional for conserved quantities. In particular these external sources are gauge fields corresponding to each symmetry. They are as follows:
\begin{itemize}
\item Let $\varepsilon_{s}^m(x)=\varepsilon_{s\m}^m(x) dx^\mu$ be the vielbeins, which can be thought of as the gauge fields corresponding to unbroken translations $\bar P_m$.  The metric tensors are then given by $g_{s\m\n}(x) = \varepsilon_{s\m}^m(x)\eta_{mn}  \varepsilon_{s\n}^n(x)$. 
\item Let $\cA_{s}(x)\equiv \cA_{s\m}^A(x) dx^\mu  T_A$ be the gauge fields (or spin connections if the Lorentz group is involved) corresponding to the unbroken zero-form symmetries other than translations.  
\item Let $c_{s}(x) \equiv c_{s\m}^\alpha(x) dx^\mu \tau_\alpha $ be the gauge fields (or spin connections) corresponding to the broken zero-form symmetries. 
\item Let $H^{p}_{s} = \frac{1}{(p+1)!} H_{s\mu_0\dots\mu_{p}}^{p}  dx^ {\mu_0}\wedge \dots\wedge dx^{\mu_{p}} \bQ_p$ be the gauge fields corresponding to the $p>0$-form symmetries.  
\end{itemize}
On each leg of the SK contour, we can combine these fields into a single object,
\begin{equation} \theta_{s}(x) = i \varepsilon_{s}^m(x)\bar P_m+ i c_{s}^\alpha(x) \tau_\alpha + i\cA_{s}^A(x) T_A + i H_s^p(x) \bQ_p, \end{equation}
where the factors of $i$ are included as a matter of convention. Now, letting $U(t,t';\theta_{s})$ for $s=1,2$ be the time-evolution operator from $t'$ to $t$ in the presence of external sources $\theta_{s}$, the generating functional for the conserved currents is
\begin{equation}e^{W[\theta_{1},\theta_{2}]} = \text{\tr}\squ{U(+\infty,-\infty; \theta_{1})\rho U^\dagger(+\infty,-\infty; \theta_{2})}. \end{equation}
Since $\theta_{s}$ couple to conserved currents, $W[\theta_{1},\theta_{2}]$ must be invariant under two independent copies of the gauge symmetries~\cite{H. Liu}; that is, for gauge parameters $\zeta_1(x)$ and $\zeta_2(x)$ we have
\begin{equation}W[\theta_{1\m},\theta_{2\m}] = W[\theta_{1\m}^{\zeta_1},\theta_{2\m}^{\zeta_2}].\end{equation} 
We can therefore `integrate in' both the broken and unbroken Goldstone fields using the Stückelberg trick. In particular, define
\begin{equation}\Theta_{s}(\sigma)\equiv \theta_s^{\zeta_s}(\sigma),~~~~~~~~~~\zeta_s\equiv \{X_s^\mu, \pi_s^\alpha,b_s^A, \varphi_s^p\},  \end{equation}
where $\sigma^M$ for $M=0,\dots,d$ are the fluid worldvolume coordinates. 
Now we can implicitly define the non-equilibrium effective action by 
\begin{equation}e^{W[\theta_{1},\theta_{2}]} \equiv \int \cD [X_s^\m  \pi_s^\alpha \epsilon_s^A\varphi_s^p] ~e^{i I_\text{EFT}[\Theta_{1},\Theta_{2}]}.\end{equation}
Notice that if we remove the external source fields by fixing $\varepsilon_{s\m}^m(x)=\delta_\m^m$ and $\cA_{s\m}^A=c_{s\m}^\alpha=H_s^p=0$, we find that $\Theta_{s}$ are nothing other than the Maurer-Cartan mixed-forms~(\ref{MC form p}) defined on each leg of the SK contour. Since $\varepsilon_{s\m}^m(x)=\delta_\m^m$ we can identify the Lorentz indices $m,n$ with the physical spacetime coordinate indices $\m,\n$, allowing us to use $\bar P_\m$ to refer to the unbroken translation generators. Moreover because these fields live on the SK contour, their values must match up in the infinite future, meaning that even though there were two copies of the gauge fields and gauge symmetries, there is only one copy of the global symmetry group $\cG$. 

Since $W[\theta_1,\theta_2]$ is the generating functional for conserved quantities, we see that any effective action constructed exclusively with building-blocks furnished by the Maurer-Cartan form can be gauged and will have conserved Noether currents associated with each of its symmetry generators.

\subsection{Dual theories and mixed 't Hooft anomalies}

We now investigate dualities among higher-form currents. Recall the example of the $3+1$-dimensional superfluid from \S\ref{Superfluids at zero temperature} in which there were two conserved charges. One was the $U(1)$ symmetry associated with particle number conservation and the other was associated with the higher-form current $K^{\mu\nu\lambda} = \epsilon^{\mu\nu\lambda\rho} \partial_\rho \psi$. Notice that $K^{\mu\nu\lambda}$ is conserved automatically as an identity. However, suppose we gauge the $U(1)$ symmetry by introducing gauge field $\cA_\mu$. This can be accomplished by replacing $\partial_\mu \psi \to \cA_\mu +\partial_\mu \psi$. Then, the higher-form current becomes $K^{\mu\nu\lambda} = \epsilon^{\mu\nu\lambda\rho}(\cA_\rho+ \partial_\rho \psi)$, which is no longer conserved; in particular, we have
\begin{equation}\partial_\mu K^{\mu\nu\lambda} = \star F ^{\nu\lambda},\end{equation}
where $F=d\cA$. Because the introduction of a gauge field for the $U(1)$ symmetry interferes with the conservation of the higher-form charge, we say that there exists a {\it mixed 't Hooft anomaly}. In particular, it is impossible to gauge the $U(1)$ symmetry and the two-form symmetry simultaneously. But recall how the arguments of the previous subsection relied on the assumption that all symmetries could be gauged simultaneously. Thus, if there is a mixed anomaly, we may only include Goldstones for one of the anomalous symmetries in the coset construction. For the example of the superfluid, we may choose to parameterize the symmetry group element with {\it either} the $U(1)$ symmetry generator $N$ and its corresponding Goldstone {\it or} the two-form generator $\bQ$ and its corresponding Goldstone, but not both. It turns out that the presence of a mixed 't Hooft anomaly is (nearly) equivalent to SSB in that it can be used to prove a kind of Goldstone theorem~\cite{Delacretaz:2019brr}.

Statements regarding mixed anomalies can be generalized to more complicated scenarios. Working in $d+1$ spacetime dimensions, suppose that $\bQ$ and $\bR$ are $p$-form and $d-p-1$-form $U(1)$ symmetry generators, respectively and let them have corresponding conserved currents $\bJ^{\mu_0\dots \mu_p}$ and $\bK^{\mu_1\dots \mu_{d-p}}$. Suppose further that $\bQ$ and $\bR$ enjoy a mixed anomaly such that if we introduce an external gauge field $H_{\mu_0\dots\mu_p}$ for $\bQ$, we have 
\begin{equation}\partial_{\mu_1}\bK^{\mu_1\dots \mu_{d-p}} = \frac{1}{(p+1)!} \epsilon^{\mu_2\dots\mu_{d-p} \nu_0\dots\nu_{p}} (d H)_{\nu_0\dots\nu_{p}} . \end{equation} 
Then, we may either include the Goldstone $\varphi$ for $\bQ$ or the Goldstone $\chi$ for $\bR$ in the Maurer-Cartan form, but not both. As we will see, the resulting effective actions will be related by a Legendre transformation in much the same way that the ordinary and dual superfluid actions are related to one another. 

Supposing that we construct our action with $\varphi$ as the Goldstone field associated with $\bQ$. Then, we have that the $d-p-1$-form current, which is identically conserved, is given by
\begin{equation} \bK = \star d \varphi ,\end{equation}
and the  $p$-form current $\bJ$, which is conserved on-shell, is obtained by a Noether procedure. Conversely, if we construct an action with $\chi$ as the Goldstone associated with $\bR$, we find that 
\begin{equation} \bJ = \star d \chi \end{equation}
is identically conserved and $\bK$, which is conserved on-shell, is obtained via a Noether procedure. Finally, if we are constructing a non-equilibrium EFT, then we must of course double the fields: $\varphi_s$ and~$\chi_s$ for $s=1,2$. 

To see explicitly how this duality works, suppose we have the non-equilibrium action $I_\text{EFT} [d\varphi_1,d\varphi_2]$, which depends on the Goldstones $\varphi_s$ associated with $\bQ$ and contains no Goldstones associated with $\bR$. This action may depend on other fields, but they will not be important for present considerations.  Then, we may replace $d\varphi_s \to \cV_s$ for generic $p+1$-forms $\cV_s$ and define an auxiliary action by
\begin{equation}\label{Legendre} I_\text{AUX} = I_\text{EFT} [\cV_1,\cV_2] - \int \cV_1\wedge d\chi_1+  \int \cV_2\wedge d\chi_2 .  \end{equation}
Notice that the equations of motion for $\chi_s$ give $d\cV_s=0$, which can be solved---assuming a topologically trivial spacetime---to give $\cV_s = d\varphi_s$ for some $p$-forms $\varphi_s$. We thus recover the original action $I_\text{EFT}$. Conversely, we may compute the equations of motion for $\chi_s$,
\begin{equation} \frac{\delta I _\text{EFT} }{\delta \cV_1} = d \chi_1, ~~~~~  \frac{\delta I_\text{EFT} }{\delta \cV_2} = -d \chi_2. \end{equation} 
These equations of motion can then be solved\footnote{At leading order in the derivative expansion, they can be solved algebraically.} for $\cV_s$ in terms of $d\chi_s$ and any other fields that may have appeared in $I_\text{EFT}$. We thus arrive at a dual action $I_\text{DUAL}[d\chi_1,d\chi_2]$ with the Goldstones $\chi_s$ associated with $\bR$ and no Goldstones associated with $\bQ$. Moreover, $I_\text{EFT}$ and $I_\text{DUAL}$ are evidently related by the Legendre transformation~\eqref{Legendre}.

\subsection{Higher-form non-Stückelberg symmetries}

To kill the conservation of the higher-form current without killing the higher-form symmetry, it is important to understand what constitutes a $p$-form symmetry. 
Recall that all $p$-form Goldstones enjoy the time-dependent symmetry~\eqref{higher form gauge}, or equivalently
\begin{equation}\varphi_s^p \to \varphi_s^p +\alpha_s^p,\end{equation}
where $\alpha_s^p$ for $s=1,2$ are closed $p$-forms. These transformations contain within them the global $p$-form symmetry transformations, but they also contain gauge redundancies. In particular, two different sets of transformation parameters $\alpha_s^p$ and $\tilde \alpha_s^p$ correspond to the same physical transformation if and only if their differences $\alpha_s^p- \tilde \alpha_s^p$ are exact for $s=1,2$. Since in the distant future, SK boundary conditions impose $\varphi_1^p(+\infty) =\varphi_2^p(+\infty)$, the $s=1$ and $s=2$ transformation parameters must contain the same topological features, meaning that $\alpha_1^p-\alpha_2^p$ must be exact. Therefore, there is only one copy of the global $p$-form symmetry and it only transforms the retarded $p$-form Goldstones. 

We find it convenient to work in synchronous gauge
\begin{equation}\varphi^p_{s 0 M_2 \dots M_p}(\sigma) = 0. \end{equation}
Then the residual symmetry of~\eqref{higher form gauge} now has $\alpha^p_1=\alpha^p_2 = \alpha^p(\sigma^I)$ for $I=1,\dots d$, which is now both closed and time-independent. 
If the higher-form symmetry is unbroken, then we also have the gauge symmetry~(\ref{higher form gauge 1}). Notice that in either case, the set of symmetries acting on these higher-form fields is time-independent. As a result, the SK boundary conditions---the requirement that the fields be continuous on the closed-time-path\footnote{If $\phi_s$ are continuous on the closed-time-path, then in the limit that $\sigma^0\to +\infty$, we have $\phi_1(+\infty) = \phi_2(+\infty)$.}---force us to use the same $\alpha$ for $s=1,2$. Thus, the advanced higher-form fields do no transform under any shift-symmetries, so we expect that there exist advanced building-blocks that contain no derivatives and are invariant under all physical symmetries. It turns out that this is so and we can compute them in the following way. Define the higher-form symmetry generalization of~\eqref{advanced form 0}, denoted by $\bG_a$, as
\begin{equation}e^{\int_{\Sigma_p }\bG_a} = g_2^{-1}(\Sigma_p) g_1(\Sigma_p). \end{equation}
Notice that $\bG_a$ is an element of the Lie algebra as its exponential is a group element. We can therefore express it as a linear combination of symmetry generators
\begin{equation}\label{advanced form}\bG_a \supset i \Phi_a^p \bQ_p +\cdots.\end{equation}
The inclusion of $\Phi_a^p$ as a building-block for the action means that $\bQ_p$ is a NSS generator. To see that this is the case, notice that in order to make $\varphi_s^p$ Stückelberg fields, we must be able to introduce external $p+1$ gauge fields $H_s^{p+1}$ such that the fields must always appear in the combinations $H_s^{p+1}+d\varphi_s^p$. But if $\Phi_a^p$ is a legitimate building-block of the action, that is if $\varphi_a^p$ may appear without derivatives, then no such external gauge fields can be introduced and $\varphi_s^p$ cannot be viewed as Stückelberg fields. 

It may seem strange that we are gauge-fixing the fields before constructing the action. Usually, one constructs the action and then gauge-fixes; after all, the components $\varphi_{s 0 M_2\dots M_p}$ yield important constraints as their equations of motion. However, since we are aiming to destroy the conservation of the $p$-form current, these constraint equations are not desirable. It is therefore appropriate---indeed necessary---to impose the synchronous gauge-fixing conditions {\it before} constructing the effective action. 

\subsection{Quadratic examples}

To illustrate some key features of mixed 't Hooft anomalies and higher-form NSS, we will investigate some simple examples. For concreteness, we will work in $3+1$ dimensions and suppose the existence of two one-form charges with mixed anomaly, $\bQ_\text{el}$ and $\bQ_\text{mag}$, which we interpret as the electric and magnetic one-form charges from electromagnetism. Working at finite temperature $1/\beta_0$, and ignoring the conservation of the energy-momentum tensor, we can fix $X_s^\mu(\sigma) = \delta_M^\mu \sigma^M$. Then we can construct our action on the physical spacetime coordinates $x^\mu \equiv X_s^\mu$. Let $A_{s\mu}$ and $\varphi_{s\mu}$ be the Goldstone fields associated with $\bQ_\text{el}$ and $\bQ_\text{mag}$, respectively. As $\bQ_\text{el}$ and $\bQ_\text{mag}$ enjoy a mixed anomaly, we do not permit both $A_s$ and $\varphi_s$ to appear in the same effective action.

In the $A$-picture, let
\begin{equation}\label{field strength E B} F_s = d A_s~~~~~~~~~~  E_s^i = F_s^{0i},~~~~~~~~~~B_s^i = \frac{1}{2}\epsilon^{ijk} F_s^{jk},\end{equation}
and in the $\varphi$-picture, let
\begin{equation}\label{field strength eb} f_s=d\varphi_s,~~~~~~~~~~{B'}_s^i = f_s^{0i},~~~~~~~~~~{E'}_s^i = \frac{1}{2}\epsilon^{ijk} f_s^{jk}.\end{equation}
We will work in the retarded-advanced basis, but for economy of expression we will drop all $r$-subscripts on retarded fields. 
All actions we construct will involve only the relevant and marginal terms and will respect the dynamical KMS symmetry. \\

\subsubsection{Maxwell in a medium} 

Begin by supposing that $\bQ_\text{el}$ and $\bQ_\text{mag}$ are both conserved and enjoy a mixed 't Hooft anomaly; as a result, they are both spontaneously broken. The fact that they are both conserved means that there are no electric or magnetic monopoles. The resulting theory will be the source-free Maxwell theory at finite temperature. Then, we can choose to work in the $\varphi$- or the $A$-picture. We choose to work in the $A$-picture as this choice will highlight the similarities with ordinary electromagnetic theory in vacuum. The action consisting of relevant and marginal terms is
\begin{equation} I = \int d^4 x \cur{\vec E_a\cdot \vec E - c_s^2 \vec B_a\cdot \vec B}. \end{equation}
The equations of motion are then
\begin{equation}\vec \nabla\cdot \vec E=0,~~~~~~~~~~\dot{\vec E} = c_s^2 \vec\nabla\times\vec B. \end{equation}
We additionally have the mathematical identities
\begin{equation}\label{Maxwell identities} \vec\nabla\cdot \vec B=0,~~~~~~~~~~\dot{\vec B} = -\vec\nabla\times \vec E. \end{equation}
Evidently, there are no electric or magnetic monopoles in this theory. Combining the above equations, we find that
\begin{equation}\ddot{\vec B} = c_s^2 \vec\nabla ^2 \vec B,~~~~~~~~~~ \ddot{\vec E} = c_s^2 \vec\nabla ^2 \vec E. \end{equation}
We thus have a wave that propagates at speed $c_s$. Notice that our equations look just like the vacuum Maxwell equations except the speed of wave-propagation is no longer the standard speed of light. This discrepancy arises because in our example, Lorentz boosts are explicitly broken. 

Now let us consider the conserved currents associated with the one-form symmetries. We have $J_\text{el}^{\mu\nu}$ and $J_\text{mag}^{\mu\nu}$, where 
\begin{equation}\label{Hot Maxwell currents} J_\text{el}^{0i} = E^i,~~~~~~~~~~J_\text{el}^{ij} = c_s^2 \epsilon^{ijk} B^k,~~~~~~~~~~ J_\text{mag}^{0i} = B^i,~~~~~~~~~~J_\text{mag}^{ij} = -\epsilon^{ijk} E^k. \end{equation}
Then, we have the conservation equations: $\partial_\mu J_\text{el}^{\mu\nu} =0$, which holds on-shell, and $\partial_\mu J_\text{mag}^{\mu\nu} =0$, which holds as an identity. Notice that expressions for the two different currents involve the same building-blocks, allowing us to express the electric higher-form current entirely in terms of components of the magnetic higher-form current, namely
\begin{equation}\label{special relationship}J_\text{el}^{0i} = \frac{c_s^2}{2} \epsilon^{ijk} J^{jk}_\text{mag},~~~~~~~~~~ J_\text{el}^{ij} = \epsilon^{ijk} J_\text{mag}^{0k}. \end{equation}
We can therefore express the conservation of the electric higher-form current in terms of the components of the magnetic higher-form current by
\begin{equation}\epsilon^{ijk} \cur{\partial_j J_\text{mag}^{0k} +\frac{c_s^2}{2} \partial_t J_\text{mag}^{jk} }=0,~~~~~~~~~~ \partial_i J_\text{mag}^{0i} = 0.\end{equation} 
Notice that if $\bQ_\text{el}$ and $\bQ_\text{mag}$ were unrelated charges, then we would have two independent conserved current, that is there would be no special relationship between the components of the two currents. In this case, we would have two diffusive modes. But in our case, there is a mixed anomaly, which relates the components of the conserved charges. This special relationship~\eqref{special relationship} along with the conservation equations is sufficient to derive the existence of a propagating wave. We therefore see the existence of wave solutions is intimately connected with the existence of a mixed anomaly. Moreover these findings readily generalize to arbitrary dimension and arbitrary $p$- and $d-p-1$-form $U(1)$ symmetries with mixed 't~Hooft anomaly. \\

\subsubsection{Magnetohydrodynamics I} 

We will now remove $\bQ_\text{el}$ as a physical conserved quantity of the theory by supposing that $\bQ_\text{mag}$ is not spontaneously broken. Physically, the resulting action will correspond to a theory of MHD in which we ignore the fluid degrees of freedom arising from the conservation of the stress-energy tensor. We work in the $\varphi$-picture and impose the time-independent shift symmetries
\begin{equation}\label{MHD gauge tran}\varphi_{si} \to \varphi_{si} + \kappa_i(\vec x) , \end{equation}
for arbitrary $\kappa_i(\vec x)$, 
which ensure $\bQ_\text{mag}$ is unbroken and that $\bQ_\text{el}$ is not gauge-invariant (see below for more details). Thus, the conservation of $\bQ_\text{el}$ is entirely removed from the theory. 
The action consisting of relevant and marginal terms is
\begin{equation}\label{MHD I} I = \int d^4 x \cur{ i \frac{D}{\beta_0} {\vec E'_a}{}^2+{\vec B'_a}\cdot {\vec B'} -D{\vec E'_a}\cdot {\dot{\vec E}'}  } , \end{equation} 
and the resulting equations of motion are
\begin{equation}\label{MHD diffusion}\vec\nabla\cdot\vec B'=0,~~~~~~~~~~\dot{\vec B}' = D\vec\nabla^2 \vec B'. \end{equation}
We therefore find a diffusion equation for the magnetic field. Compare this result with that of the previous example: when $\bQ_\text{mag}$ is spontaneously broken, there exists a propagating wave; when it is unbroken, there is a diffusive mode. This relationship between SSB pattern and dispersion relation in non-equilibrium systems is quite common~\cite{Landry}.

Now consider the conserved currents $j_\text{el}^{\mu\nu}$ and $j_\text{mag}^{\mu\nu}$ associated with $\bQ_\text{el}$ and $\bQ_\text{mag}$, respectively.\footnote{We used calpital letter $J$ when working in the $A$-picture and lower-case $j$ when working in the $\varphi$-picture.} Explicitly, these currents are
\begin{equation}\label{MHD I currents} j_\text{el}^{0i} = {E'}^i,~~~~~~~~~~j_\text{el}^{ij} = \epsilon^{ijk} {B'}^k,~~~~~~~~~~ j_\text{mag}^{0i} = {B'}^i,~~~~~~~~~~j_\text{mag}^{ij} =D \epsilon^{ijk} \dot {E}'{}^k. \end{equation}
Notice that $\partial_\mu j_\text{mag}^{\mu\nu}=0$ on the equations of motion and $\partial_\mu j_\text{el}^{\mu\nu}=0$ identically. It therefore seems like there are two conserved one-form symmetries; however $j_\text{el}^{\mu\nu}$ is not invariant under the gauge transformation~\eqref{MHD gauge tran} and is hence not physical. We therefore only have one conserved current, namely $j_\text{mag}$. 
\\

\subsubsection{Magnetohydrodynamics II} 

We will now consider a gentler way of removing $\bQ_\text{el}$ from the set of conserved charges. We will see that we can make the non-conservation of this charge arbitrarily weak by adjusting a time-scale~$\tau$. To do this, work in the $A$-picture and let $\bQ_\text{el}$ be a spontaneously broken NSS charge. As a result, $\bQ_\text{mag}$ will still exist, but it will no longer enjoy a mixed anomaly with another conserved charge, which indicates that it is {\it not} spontaneously broken. Thus, if we only consider conserved charges, we have the same SSB pattern as in the previous example. The resulting theory will therefore describe MHD, but unlike the more standard formalism, this new theory allows for large charge density.

In order for $\bQ_\text{el}$ to be the generator of a NSS, we must operate in synchronous gauge, that is we fix $A_{s0} = 0$, which yields the residual symmetry
\begin{equation}A_s \to A_s + \alpha(\vec x),\end{equation} 
for closed, time-independent one-form $\alpha$ such that $\alpha_{0}=0$. Notice that $\vec A_a$ is now a covariant building-block. The action consisting of relevant and marginal terms is therefore 
\begin{equation}\label{MHD II} I =\int d^4 x\cur{\vec E_a\cdot \vec E - c_s^2 \vec B_a\cdot \vec B + \frac{1}{\tau} \vec A_a \cdot \vec E +\frac{i}{\tau\beta_0} \vec A_a^2 }.\end{equation}
The terms with a factor of $1/\tau$ result from the symmetry being a NSS. We can make this relaxation time $\tau$ arbitrarily long.

The  resulting equations of motion are
\begin{equation} \dot {\vec E} - c_s^2 \vec\nabla\times\vec B = - \frac{1}{\tau} \vec E . \end{equation} 
Since we have no $\vec A_{s0}$-field in the action, the electric Gauss law is not an equation of motion. 
Taking the curl and divergence, respectively, of the above equation yields
\begin{equation} \ddot {\vec B} -c_s^2 \vec\nabla^2 \vec B  = -\frac{1}{\tau} \dot{\vec B},~~~~~~~~~~\cur{\partial_t +\frac{1}{\tau} } \vec\nabla\cdot \vec E = 0,\end{equation} 
where we have made use of the identities~\eqref{Maxwell identities}. 
Using the Maxwell relation $\vec\nabla \cdot \vec E = \rho$, where $\rho$ is the electric charge density, we see that the second equation can be solved by $\rho(t,\vec x) = \rho_0(\vec x) e^{-t/\tau}$, meaning that non-zero charge density can exist, but it decays exponentially fast.

Let us now investigate the connection to MHD and free Maxwell theory. To recover the MHD equations of motion, work in the low-frequency limit $\partial_t \ll 1/\tau$. Then, the equations of motion become
\begin{equation}\label{MHD II eom 1} \dot{\vec B} = \tau c_s^2 \vec\nabla^2 \vec B ,~~~~~~~~~~\rho = 0.\end{equation} 
If we identify $D=\tau c_s^2$, then we recover the diffusive MHD equation~\eqref{MHD diffusion}. Further, notice that  $\rho=0$, indicating that there can be no electric charge density. These equations of motion along with the identities~\eqref{Maxwell identities} are exactly what we expect for standard MHD. To see the connection with free Maxwell theory, work in the high-frequency limit $\partial_t\gg 1/\tau$. Then, we have
\begin{equation}\label{MHD II eom 2} \ddot{\vec B} = c_s^2 \vec\nabla^2 \vec B,~~~~~~~~~~ \partial_t \rho = 0.  \end{equation}
The first equation indicates that there now exists a propagating wave and the second equation indicates that we are operating on sufficiently short time scales that charge density does not have time to decay.\footnote{If we included the conservation of the stress-energy tensor, the second equation of~\eqref{MHD II eom 2} would tell us that the charge in each fluid element is `locked in' at leading order in the derivative expansion. If higher-order corrections are considered, this equation becomes diffusive.} 
If we take the initial condition $\rho_{t=0}=0$, then the the high-frequency equations of motion in conjunction with the identities~\eqref{Maxwell identities} reproduce the free Maxwell equations except the speed of light is replaced with $c_s$. 

We arrive at a fascinating result: on long time-scales, $\bQ_\text{mag}$ appears unbroken, while on short time-scales, it appears spontaneously broken. Moreover, the characteristic time scale, $\tau$ at which this crossover occurs is fixed by the strength of the current non-conservation. We expect this scale-dependent symmetry-breaking to be a much more general phenomenon. 

Finally, the higher-form currents are given by~\eqref{Hot Maxwell currents}, but now only one of them is conserved. In particular, $J_\text{mag}$ is conserved as an identity, while 
\begin{equation} \partial_\mu J_\text{el}^{\mu i } =-\frac{1}{\tau} J_\text{el}^{0i} .\end{equation}
We see that on time-scales much longer than $\tau$, the r.h.s. of the above equation dominates and we only have one conserved current, just as in the standard MHD example. By contrast, on time-scales much shorter than $\tau$, the l.h.s. of the above equation dominates and $J_\text{el}$ appears essentially conserved, which resembles the Maxwell in a medium example. Notice that there still exists a relationship among the components of the two higher-form currents~\eqref{special relationship} even though $\bQ_\text{el}$ is no longer conserved. We therefore see that in some sense the mixed 't Hooft anomaly still exists. 
 \\

\subsubsection{Gapped diffusion} 

We now consider the situation in which both $\bQ_\text{el}$ and $\bQ_{mag}$ are not conserved. Working in the $\varphi$-picture, we take $\bQ_\text{mag}$ to be an unbroken charge associated with a NSS. Therefore we must impose synchronous gauge ($\varphi_{s0}=0$) and endow $\varphi_s$ with the time-independent gauge symmetries
\begin{equation}\varphi_{si} \to \varphi_{si} +\kappa_i(\vec x) , \end{equation}
for arbitrary $\kappa_i(\vec x)$. Notice that $\varphi_{ai}$ is an allowed building-block. The action consisting of relevant and marginal terms is then
\begin{equation} I = \int d^4 x \cur{ i \frac{D}{\beta_0} {\vec E'_a}{}^2+{\vec B'_a}\cdot {\vec B'} -D{\vec E'_a}\cdot {\dot{\vec E}'} +\frac{1}{\tau} \vec \varphi_a\cdot \vec B' +\frac{i}{\tau\beta_0} \vec\varphi_a^2  } . \end{equation} 
The equations of motion are therefore 
\begin{equation} \cur{\partial_t +\frac{1}{\tau} } \vec B' = D\vec\nabla^2 \vec B', \end{equation}
which is a gapped diffusion equation. Thus, there are no gapless excitations.

The higher-form currents are given by~\eqref{MHD I currents}, such that $j_\text{el}$ is not gauge invariant and is therefore not a physical current. But now, $j_\text{mag}$ is not conserved, namely
\begin{equation}\partial_\mu j_\text{mag}^{\mu i} =- \frac{1}{\tau} j_\text{mag}^{0 i} . \end{equation} 
Thus, we see explicitly that there are no conserved currents in this theory, which is why there are no gapless excitation.

\subsubsection{Legendre transform for unbroken symmetries}

We have seen that when a $p$- and $d-p-1$-form current are conserved and enjoy a mixed anomaly, it is possible to perform a Legendre transform to convert the theory involving the $p$-form Goldstone into a dual theory involving the $d-p-1$-form Goldstone. Because these actions are related by a Legendre transform, we can confidently say that they are physically equivalent. By contrast, we merely provided general arguments that the two distinct actions for MHD given by~\eqref{MHD I} and~\eqref{MHD II} were in some sense physically equivalent. Here we will show that there is a duality relating these two actions. 

Begin by considering~\eqref{MHD I} and replace $\vec E' _s \to \vec e_s$ and $\vec B'_s \to \vec b_s$, where $\vec e_s$ and $\vec b_s$ are fundamental fields. We will work in the retarded-advanced basis and drop the $r$-subscript on the retarded fields. Notice that $\vec e$ always appears with a time-derivative, meaning that there is a gauge symmetry 
\begin{equation}\label{gauge sym MHD legendre} \vec e\to \vec e+\vec \xi(\vec x),\end{equation}
for arbitrary time-independent $\vec \xi$. 
Following~\eqref{Legendre}, we would like to construct the auxiliary action 
 \begin{equation} I_\text{AUX}  \overset ? = \int d^4 x \cur{ i \frac{D}{\beta_0} {\vec e_a}{}^2+{\vec b_a}\cdot {\vec b} -D{\vec e_a}\cdot {\dot{\vec e}}   + \vec e\cdot \vec E_a +\vec e_a\cdot \vec E-\vec b\cdot \vec B_a-\vec b_a\cdot \vec B}, \end{equation} 
where $\vec E_s$ and $\vec B_s$ are given by~\eqref{field strength E B}. Unfortunately this action contains the term 
\begin{equation}\vec e \cdot \vec E_a = \vec e \cdot \cur{\vec\nabla A_{a0} -\partial_t \vec A_a },\end{equation}
which is not gauge invariant. If, however, we impose $A_{a0}=0$, then integration by parts removes the gauge non-invariant term. We are then left with the auxiliary action
 \begin{equation}\label{AUX MHD} I_\text{AUX}  = \int d^4 x \cur{ i \frac{D}{\beta_0} {\vec e_a}{}^2+{\vec b_a}\cdot {\vec b} -D{\vec e_a}\cdot {\dot{\vec e}}   +\dot{ \vec e}\cdot  {\vec A}_a +{\vec e}_a\cdot  {\vec E}-\vec b\cdot \vec B_a-\vec b_a\cdot \vec B}. \end{equation} 
Notice that the equations of motion for $\vec A_s$ and $A_{0}$ yield
\begin{equation} \vec\nabla\times\vec b_s = -\dot{\vec e}_s,~~~~~~~~~~\vec\nabla\cdot \vec e_a=0,\end{equation} 
which, with appropriate gauge choice for~\eqref{gauge sym MHD legendre}, imply that there exist one-forms $\varphi_s$ such that 
\begin{equation} \vec e_s = \vec E'_s,~~~~~~~~~~\vec b_s = \vec B'_s,\end{equation}
where $ \vec E'_s$ and $ \vec B'_s$ are given by~\eqref{field strength eb}.  Plugging these solutions into $I_\text{AUX}$, we recover the original action~\eqref{MHD I}. 

Conversely, to find the dual action, we may instead integrate out $\vec e_s$ and $\vec b_s$.  The equations for $\vec b_s$ are very straight-forward, 
\begin{equation}\label{magnetic eom} \vec b_s = \vec B_s. \end{equation}
 The equations for $\vec e_s$ are
\begin{equation}\label{electric eom} \frac{2Di}{\beta_0} \vec e_a - D\dot{\vec e} + {\vec E} = 0,~~~~~~~~~~\partial_t \cur{D{\vec e}_a-\vec A_a} = 0.   \end{equation}
The second equation can be integrated over time to give $D \vec e_a +\vec \lambda (\vec x) =  \vec A_a,$ 
where $\vec\lambda$ is an arbitrary integration function. But the SK boundary condition mandates that all advanced fields vanish in the infinite future, forcing $\vec \lambda = 0$, which yields
\begin{equation} \label{electric a eom} \vec e_a =\frac{1}{D} \vec A_a. \end{equation}
And finally, the first equation of~\eqref{electric eom} can be solved to give
\begin{equation} \label{electric r eom} \dot {\vec e} = \frac{2i}{D\beta_0} \vec A_a+\frac{1}{D} \vec E.  \end{equation}
Plugging the solutions~\eqref{magnetic eom}, \eqref{electric a eom}, and~\eqref{electric r eom} into the auxiliary action~\eqref{AUX MHD}, we obtain the dual action
\begin{equation}\label{MHD IIb} I_\text{DUAL} =\int d^4 x\cur{- c_s^2 \vec B_a\cdot \vec B + \frac{1}{\tau} \vec A_a \cdot \vec E +\frac{i}{\tau\beta_0} \vec A_a^2 },\end{equation}
where we have identified $D=\tau c_s^2$ and re-scaled the fields by $\vec A_s\to c_s \vec A_s$. At this stage we may gauge-fix $A_{0}=0$ with no consequence as the resulting equations of motion from this component will be entirely redundant. Notice that this dual action exactly matches~\eqref{MHD II} except for the term $\vec E_a\cdot \vec E$. The reason this term is missing has to do with power-counting: in the original action~\eqref{MHD I}, the dynamics are diffusive, so we consider $\partial_t\sim \vec\nabla^2$. If $\tau$ is large, that is $1/\tau$ is of order the energy cutoff for the EFT, then we should expect $\partial_t\sim \vec\nabla^2$ in the dual action~\eqref{MHD IIb}. In such a case, the term $\vec E_a\cdot \vec E$ must be considered irrelevant, so we ought not include it in the leading-order action. By contrast, when we constructed~\eqref{MHD II} from symmetry considerations, we assumed that the non-conservation of the NSS was weak, meaning that $1/\tau$ is much less than the cutoff energy. As a result, we may consider $\partial_t\sim \vec\nabla$, making the term $\vec E_a\cdot \vec E$ marginal.

\section{Dual superfluids}

In this section, we will construct the dual two-form EFT for $3+1$-dimensional superfluids at zero and finite temperature. For a coset construction of the ordinary superfluid action, consult~\cite{Landry}. 

\subsection{Zero temperature}

In the dual description of the superfluid, we have, in addition to Poincaré symmetry, the $U(1)$ two-form symmetry generator $\bQ$ and $U(1)$ zero-form generator $N$, which counts particle number. Moreover, there is a mixed anomaly between $\bQ$ and $N$, meaning that we can only include one of them in our coset construction.

First, suppose we take $N$ to be in our coset. Then, the standard claim is that both $Q$ and $P_0$ are spontaneously broken, but their diagonal subgroup $\bar P_0 = P_0 +\mu N$ is preserved~\cite{Son,Witten}. Boosts are also spontaneously broken but all other symmetries are preserved. Thus the most general coset element takes the form
\begin{equation} \label{broken P0}g(x) = e^{i x^\mu \bar P_\mu} e^{i \pi(x) N} e^{i\eta^i(x) K_i} .  \end{equation}
Notice that we can rewrite this group element as
\begin{equation}\label{unbroken P0} g(x) = e^{i x^\mu P_\mu} e^{i \psi(x) N} e^{i\eta^i(x) K_i} ,~~~~~~~~~~\psi\equiv \mu t + \pi.  \end{equation}
Thus, we can forget about the fact that $P_0$ is broken; instead we may act as though only $N$ is broken and then after the fact give $\psi$ a time-dependent VEV. The coset construction for the ordinary superfluid action is already well-studied, so we will now change gears and consider the dual theory.

In the dual picture, we take Lorentz boosts and $\bQ$ to be spontaneously broken and we suppose our system exists at finite particle-number density. Because we are constructing an EFT describing a zero-temperature system, we need not bother with the SK formalism, so there is just one copy of each field. Further, we define our coset on the physical spacetime directly and we only include Goldstones associated with broken symmetries. The most general coset element is 
\begin{equation} g(\Sigma) = e^{i x^\mu P_\mu} e^{i\eta^i(x) K_i} e^{i \bQ \int_{\Sigma_2}A },  \end{equation}
where $\Sigma = \{x,\Sigma_2\}$, for two-dimensional manifold $\Sigma_2$, and where $A$ is a two-form field. Notice that in the dual picture, we do not treat $P_0$ as spontaneously broken. Much like~\eqref{unbroken P0} we must instead give $A(x)$ a spacetime-dependent VEV. It turns out that requiring 
\begin{equation}\label{VEV superfluid} A_{\mu\nu} = \vev{A_{\mu\nu}} + \varphi_{\mu\nu},~~~~~~~~~~  \vev{A_{0i}} = 0,~ \vev{A_{ij}} \propto \epsilon_{ijk} x^k, \end{equation}
is equivalent to having a dual theory at finite particle-number density~\cite{Horn:2015zna}. Thus, it is the dual-theory equivalent of having $P_0$ and $N$ broken but their diagonal subgroup preserved in the original picture~\eqref{broken P0}. 

The resulting Maurer-Cartan form is
\begin{equation}\Omega = i E^\nu(P_\nu + \nabla_\nu \eta^i )+ \omega^i J_i+\star F \bQ,\end{equation}
where
\begin{equation}\begin{split}
E_\mu^\nu &= {\Lambda^\nu}_\mu, \\
\nabla_\mu\eta^i&=(E^{-1})_\mu^\nu [\Lambda^{-1} \partial_\nu\Lambda]^{0i},\\
\omega_\mu^i & = \frac{1}{2} \epsilon^{ijk }[\Lambda^{-1} \partial_\mu \Lambda]^{jk},\\
F^\mu & = \frac{1}{2} \epsilon^{\mu\nu\lambda\rho} \partial_\nu A_{\lambda\rho}. 
\end{split}\end{equation}
It is convenient to also define $\cF^\mu = F^\nu E_\nu^\mu$, which is boost-invariant.  

Now impose IH constraints to remove the boost Goldstones. In particular, fix
\begin{equation}\label{superfluid IH} 0 = \cF^i.  \end{equation} 
The quickest way to solve this equation is to note that $e_\mu^{(\nu)} = {\Lambda_\mu}^\nu$ form a set of orthonormal vectors such that $e_\mu^{(\nu)} \eta^{\mu\mu'} e_{\mu'}^{(\nu')} = \eta^{\nu\nu'}$. Thus, the IH constraints imply that $F^\mu$ is orthogonal to $e_\mu^{(i)}$ and hence parallel to $e_\mu^{(0)} = {\Lambda_\mu}^0$. We therefore have that
\begin{equation}{\Lambda_\mu}^0 = \frac{F_\mu}{F},\end{equation} 
where $F=\sqrt{-F^\mu F_\mu}$. In terms of the boost Goldstones $\eta^i$, we have 
\begin{equation} \frac{\eta^i}{\eta} \tanh \eta = \frac{F^i}{F^0},\end{equation}
where $\eta = \sqrt{\vec\eta^2}$ and we have used the fact that ${\Lambda_0}^0 = \cosh \eta$ and ${\Lambda_i}^0 = (\eta^i/\eta) \sinh \eta$. Notice that this equation is only non-singular because we supposed that $A_{\mu\nu}$ enjoys the VEV~\eqref{VEV superfluid}. 

With these IH constraints solved, we have just one invariant building-block at leading order in the derivative expansion, namely 
\begin{equation} Y\equiv \cF^t = \sqrt{-F^\mu F_\mu}. \end{equation}
The leading-order effective action is therefore
\begin{equation} S = \int d^4 x\, L(Y), \end{equation} 
for some function $L$. 

Notice that the zero-form charge $N$ associated with particle number conservation did not appear anywhere in this coset construction.\footnote{Similarly in the ordinary coset construction, the two-form symmetry does not appear.} The reason is that in the dual picture, the conserved current appears as 
\begin{equation} J_{U(1)}^\mu = \frac{1}{2} \epsilon^{\mu\nu\lambda\rho} \partial_\nu A_{\lambda\rho} ,\end{equation}
which is conserved off-shell as a mathematical identity.

\subsection{Finite temperature}

We now turn our attention to the finite temperature case. The SSB patter is of course the same. The difference is that now we define the action on the fluid worldvolume and parameterize the full symmetry group with Goldstones even if they are associated with unbroken symmetries. At finite temperature, we define our action on the SK contour and as a result must have doubled field content. If, however, we are only interested in the leading-order action, then it turns out that the dynamical KMS symmetries force the non-equilibrium action to factorize as the difference of two ordinary actions, each with a single copy of the fields~\cite{Landry}. We will therefore work with just one copy of the fields. Higher-order corrections require two copies of the fields. 

The most general group element is
\begin{equation} g(\Sigma) = e^{i X^\mu(\sigma) P_\mu} e^{i\eta^i(\sigma) K_i} e^{i \theta^i(\sigma) J_i} e^{i \bQ\int _{\Sigma_2} A},  \end{equation}
where $\Sigma = \{\sigma,\Sigma_2\}$, for two-dimensional manifold $\Sigma_2$. 
Then the Maurer-Cartan form is 
\begin{equation}\begin{split} \Omega = iE^\mu(P_\mu +\nabla_\mu \eta^i K_i) +i \omega^i J_i + i\star F\bQ, \end{split}\end{equation}
where 
\begin{equation}\begin{split}
E_M^\mu & = \partial_M X^\nu [\Lambda R{]_\nu}^\mu ,\\
\nabla_\mu \eta^i & = (E^{-1})_\mu^M[\Lambda^{-1}\partial_M \Lambda ]^{0j} R^{ji}, \\
\omega_M^i & = \frac{1}{2}  \epsilon^{ijk} [(\Lambda R)^{-1} \partial_M (\Lambda R)]^{jk}, \\
F^{M} & = \frac{1}{2} \epsilon^{MNPQ}\partial_N A_{PQ},
\end{split}\end{equation}
such that ${\Lambda^\mu}_\nu = (e^{i\eta^i K_i}{)^\mu}_\nu$ and $R^{ij} = (e^{i\theta^i J_i})^{ij}$. 

We can now impose IH constraints to remove the Lorentz Goldstones. First, to remove boost Goldstones, we impose~\eqref{boost IH}, which can be solved to yield 
\begin{equation} \frac{\eta^i}{\eta}\tanh \eta = -\frac{\partial_0 X^i}{\partial_0 X^t}, \end{equation}
where $\eta \equiv \sqrt{\vec\eta^2}$.\footnote{We could have alternatively imposed IH constraints $F^M (E^{-1})_M^i = 0$, in keeping with~(\ref{superfluid IH}). At least at leading-order these two IH constraints end up yielding the same EFT. } Second, to remove the rotation Goldstones, fix the constraint~\eqref{rotation IH}. As mentioned previously, we need not solve this IH constraint for leading-order actions. 

After imposing these IH constraints, the leading-order building-blocks are as follows. Begin by defining the fluid worldvolume metric by~\eqref{WV metric}. 
Then, we have the local temperature $T$ given by~\eqref{local temperature} as an invariant building-block.  
 Finally, we have that
\begin{equation} Y = \sqrt{F^M G_{MN} F^N},~~~~~~~~~ Z = F^M G_{M0}, \end{equation}
are the remaining invariant building-blocks. Thus, transforming to the physical spacetime coordinates $x^\mu = X^\mu$, the leading-order action is
\begin{equation}S = \int d^4 x \, L (T,Y,Z) . \end{equation}

\section{Dual solids}

In this section, we will construct the dual two-form EFTs for solids at zero and finite temperature. The two form charges count the number of crystal defects of the lattice. 

\subsection{Zero temperature}

At zero temperature, the standard EFT for a solid involves three Goldstone modes $\phi^i$ for $i=1,2,3$, corresponding to spontaneously broken translations~\cite{Zoology,Landry,Armas:2020bmo,Armas:2019sbe}. At leading order in the derivative expansion, we have
\begin{equation}
S = \int d^4 x\, P(\gamma^{ij}),~~~~~ \gamma^{ij} = \partial_\mu \phi^i \partial^\mu \phi^j,
\end{equation}
for some function $P$. Note that these Goldstones are invariant under the shift symmetries
\begin{equation}\phi^i\to\phi^i+c^i, \end{equation}
for constant $c^i$. Thus this effective field theory has three internal $U(1)$ symmetries, which have the interpretation of lattice momentum. Let $\bP_i$ be the generators of these $U(1)$ symmetries. We additionally have three distinct two-form symmetries with conserved currents
\begin{equation}\vec\bJ^{\mu\nu\lambda} = \epsilon^{\mu\nu\lambda\rho} \partial_\rho \vec \phi. \end{equation}
It is the aim of this subsection to take these two-form symmetries as the starting point.

Recall that the SSB pattern of solids is as follows. Boosts and rotations are spontaneously broken. Physical spatial translations and lattice-momentum translations are spontaneously broken, but their diagonal subgroups $P_i+\bP_i$ are preserved, and temporal translations are preserved. In the dual picture, we do not make use of the lattice momentum generators $\bP_i$ and hence we cannot treat them or $P_i$ as spontaneously broken. We must, however, still continue to treat boosts and rotations as spontaneously broken. 

In the dual picture, in addition to Poincaré symmetry, we have a three-vector of spontaneously broken two-form symmetries with generator $\vec \bQ$ that enjoy mixed anomalies with $\vec\bP$. As a result, if we include Goldstones for $\vec\bQ$, we cannot have Goldstones for $\vec\bP$ in the coset. 
Since we are working at zero temperature, we only need one copy of the fields, we can formulate our EFT directly on the physical spacetime manifold, and we need only include Goldstones for spontaneously broken symmetries. The most general coset element is 
\begin{equation} g(\Sigma) = e^{i x^\mu P_\mu} e^{i\eta^i(x) K_i} e^{i\theta^i(x) J_i} e^{{i}\vec\bQ \cdot \int_{\Sigma_2} \vec\varphi },  \end{equation}
where $\Sigma = \{x,\Sigma_2\}$, for two-dimensional manifold $\Sigma_2$. 
Then the resulting Maurer-Cartan form is
\begin{equation}\Omega = i  E^\nu(P_\nu + \nabla_\nu \eta^i+\nabla_\nu\theta^i J_i + \star \vec F \cdot \vec \bQ),\end{equation}
where
\begin{equation}\begin{split}
E_\mu^\nu &= {(\Lambda R)^\nu}_\mu, \\
\nabla_\mu\eta^i&=(E^{-1})_\mu^\nu [(\Lambda R)^{-1} \partial_\nu(\Lambda R)]^{0i},\\
\nabla_\mu\theta^i & = \frac{1}{2} \epsilon^{ijk } (E^{-1})_\mu^\nu[(\Lambda R)^{-1} \partial_\nu (\Lambda R)]^{jk},\\
\vec \cF^\mu & = E^\mu_\nu \vec F^\nu,
\end{split}\end{equation}
such that $\vec F^\mu\equiv \frac{1}{2} \epsilon^{\mu\nu\lambda\rho} \partial_\nu \vec \varphi_{\lambda\rho}$, ${\Lambda^\mu}_\nu = (e^{i\eta^i K_i}{)^\mu}_\nu$, and $R^{ij} = (e^{i\theta^i J_i})^{ij}$. 

We can now impose IH constraints to remove the Lorentz Goldstones. To remove the rotation Goldstones, we impose\footnote{The indices $i,j=1,2,3$ are now playing two roles: the first is the role of spatial rotation index and the second is to enumerate the components of $\vec\bQ$. We can get away with this redundant notation because rotations are spontaneously broken.} 
\begin{equation} \epsilon^{ijk} \cF^{jk} = 0, \end{equation}
which can be solved to give 
\begin{equation}\label{rotation IH solution} \cF^{ij} = (Y^{1/2})^{ij},~~~~~ Y^{ij} = F^{i\mu} F^j_\mu, \end{equation}
where $Y^{1/2}$ is the matrix square root of $Y$. 
Then, to remove the boost Goldstones, we impose
\begin{equation} \cF^{it} = 0, \end{equation}
which  can be solved to give 
\begin{equation}\label{boost IH solution} \frac{\eta^i}{\eta}\tanh \eta = -  \cF^{jt}(Y^{-1/2})^{ji}. \end{equation}
We are thus left with only one set of invariant building-blocks, namely $Y^{ij}$. The leading-order dual action is then
\begin{equation}  S_\text{DUAL} = \int d^4 x \, L (Y^{ij}), \end{equation}
for some function $L$. Notice how both the standard solid action and the dual solid action depend on symmetric $3\times 3$ matrices. Finally, as in the superfluid case, we should expect the two-form fields to have non-trivial background values. In particular, we want $F^{i\mu}\propto \eta^{i\nu}$. We therefore suppose that
\begin{equation}\label{solid VEV}\vev{\varphi^i_{0j}} = 0 ,~~~~~~~~~~\vev{\varphi^i_{jk}} \propto \epsilon^{ijk} t .\end{equation} 

\subsection{Finite temperature}

At finite temperature, the symmetry-breaking pattern is unchanged; however, lattice momentum generated by $\bP_i$ need not be conserved. If it is conserved, we have the phenomenon of second sound; otherwise there is no second sound~\cite{Landry second sound,solid second sound ref}.

Since our EFTs now exist at finite temperature, we have two copies of the fields, in particular, we have two copies of the solid fields $\phi_s^i$ for $s=1,2$. Alternatively, we have in the retarded-advanced basis, $\phi_r^i$ and $\phi_a^i$. If lattice momentum is conserved, then both $\phi_r^i$ and $\phi^i_a$ must always appear in the action with at least one derivative, namely $\partial_\mu \phi_r^i$ and $\partial_\mu \phi_a^i$. If, however, lattice momentum is not conserved, but $\bP_i$ is still a symmetry, then $\phi_a^i$ may appear without derivatives even as $\phi_r^i$ must always appear with a derivative. Recall that $\phi_r^i$ are the classical fields, while $\phi_a^i$ encode information about fluctuations. Thus, whether or not lattice momentum is conserved, we have the three-form conserved currents
\begin{equation}\vec \bJ^{\mu\nu\lambda\rho} = \epsilon^{\mu\nu\lambda\rho} \partial_\rho \vec\phi_r . \end{equation} 
If we then switch to the dual picture and work with two-form fields $\vec \varphi_{s\mu\nu},$\footnote{We can take $\vec\varphi_{s\mu\nu}$ to be the pullback of $\vec \varphi_{sMN}$ to the physical spacetime via the inverse of the map $X_s^\mu(\sigma)$.} the currents associated with $\bP_i$ are now given by 
\begin{equation} \vec J^\mu = \frac{1}{2} \epsilon^{\mu\nu\lambda\rho} \partial_\nu\vec\varphi_{r\lambda\rho}. \end{equation}
But notice that $\partial_\mu  \vec J^\mu = 0$ identically because partial derivatives commute. It therefore seems that the dual picture automatically has conserved lattice momentum; however, this need not be true if $J^\mu$ is not gauge invariant. Notice that if $\vec\bQ$ are unbroken, then we have the gauge symmetries of the form~(\ref{higher form gauge 1}), which do not leave $J^\mu$ invariant. Thus, from the dual perspective, we have conserved lattice momentum and hence second sound if and only if $\vec \bQ$ are spontaneously broken. 
\\

We now proceed to constructing the dual actions for solids with and without second sound. As in the zero-temperature case, we have spontaneously broken boosts and rotations, but we do not treat translations as broken.\footnote{Instead of treating spatial translations as spontaneously broken, we endow the two-form fields with the VEVs~\eqref{solid VEV}} We also have the three-form symmetry charges $\vec \bQ$. The most general group element is 
\begin{equation} g(\Sigma) = e^{i X^\mu(\sigma) P_\mu} e^{i\eta^i(\sigma) K_i} e^{i \theta^i(\sigma) J_i} e^{i\vec\bQ\cdot\int_{\Sigma_2} \vec\varphi},  \end{equation}
where $\Sigma = \{\sigma,\Sigma_2\}$, for two-dimensional manifold $\Sigma_2$.
And the Maurer-Cartan form is 
\begin{equation}\begin{split} \Omega = i E^\mu(P_\mu +\nabla_\mu \eta^i K_i+\nabla_\mu \theta^i J_i) + i \star \vec F\cdot\vec\bQ, \end{split}\end{equation}
where 
\begin{equation}\begin{split}
E_M^\mu & = \partial_M X^\nu [\Lambda R{]_\nu}^\mu ,\\
\nabla_\mu \eta^i & = (E^{-1})_\mu^M[\Lambda^{-1}\partial_M \Lambda ]^{0j} R^{ji}, \\
\nabla_\mu \theta^i & = \frac{1}{2}  \epsilon^{ijk}(E^{-1})_\mu^M [(\Lambda R)^{-1} \partial_M (\Lambda R)]^{jk}, \\
\vec F^{M} & = \frac{1}{2} \epsilon^{MNPQ}\partial_N \vec \varphi_{PQ},
\end{split}\end{equation}
such that ${\Lambda^\mu}_\nu = (e^{i\eta^i K_i}{)^\mu}_\nu$ and $R^{ij} = (e^{i\theta^i J_i})^{ij}$. It will be convenient to define $\cF^{i\mu} = E^\mu_M F^{iM}$. 

The IH constraints necessary to remove boost Goldstones are 
\begin{equation}E_0^i=0\implies \frac{\eta^i}{\eta}\tanh \eta = -\frac{\partial_0 X^i}{\partial_0 X^t},\end{equation}
and the IH constraints necessary to remove rotation Goldstones are
\begin{equation} \epsilon^{ijk} \cF^{jk}=0 \implies \cF^{ij} = (y^{1/2})^{ij},\end{equation}
where $y^{ji} = \cF^{i\mu}\cF^j_\mu$. 

The invariant building-blocks are as follows. First, we have the local temperature given by~(\ref{local temperature}). Next, we have the symmetric matrix $y^{ij}$ as defined above; notice that both of these terms are invariant under the gauge symmetry 
\begin{equation}\label{xxx}\varphi_{IJ}(\sigma) \to \varphi_{IJ}(\sigma) + \kappa_{IJ}(\sigma^K). \end{equation}
Thus these building-blocks are invariant whether or not $\vec\bQ$ are spontaneously broken. Lastly, we have the building-block
\begin{equation}z^i \equiv \cF^{i t} = u^\mu F^i_\mu,~~~~~~~~~~ 
\text{where } u^\mu \equiv T \frac{\partial X^\mu}{\partial \sigma^0},\end{equation}
but it is not invariant under the gauge symmetry~(\ref{xxx}). Thus it is only an allowed building-block if $\vec\bQ$ are spontaneously broken, that is, when lattice momentum is conserved. 

Transforming to physical spacetime, the dual actions for solids are therefore 
\begin{equation}S_\text{DUAL} = \int d^4x\,  L (T,y^{ij},z^i),\end{equation}
describing a solid with conserved lattice momentum, and 
\begin{equation}\label{no ss action}S_\text{DUAL} = \int d^4x \, L (T,y^{ij}),\end{equation}
describing a solid without conserved lattice momentum. 
The building-block $z^i$ captures the relative motion of the fluid\footnote{Second sound emerges when a thermal cloud of solid phonons can flow as a fluid independently of the solid lattice~\cite{solid second sound ref,Nature second sound}. Then we have both fluid sound waves and solid (i.e. lattice) sound waves.} and the lattice when second sound is present.

\section{Discussion}

In this work, we presented a systematic procedure to formulate non-equilibrium effective actions for hydrodynamic  (gapless) and quasi-hydrodynamic (weakly-gapped) excitations involving $p$-form symmetries. To aid in the systematization, we employed the coset construction, generalizing it to account for higher-form symmetries in non-equilibrium systems. In addition, we extended the coset construction to account for theories involving non-Stückelberg symmetries, which are exact symmetries of the action but have no conserved (non-trivial) Noether current on shell. This kind of non-conservation can only occur in non-equilibrium systems.

Further, we noticed some interesting features of non-equilibrium EFTs that involve higher-form symmetries:
\begin{itemize}
\item When implementing the coset construction, the Maurer-Cartan form associated with a $p$-form symmetry, which we denote by $\Omega_{p+1}$, is a $p+1$-form. If various higher-form symmetries of different $p$ exist in a system, then the resulting Maurer-Cartan form is a mixed-form in the sense that it can be expressed as a sum of the form $\Omega = \sum_p \Omega_{p+1}$. 
\item Whenever a $U(1)$  $p$-form symmetry is spontaneously broken, there exists a $d-p-1$-form $U(1)$ symmetry that is also spontaneously broken. Further, the $p$-form and $d-p-1$-form symmetries exhibit a mixed 't Hooft anomaly. 
\item Whenever such a mixed anomaly exists, it is impossible to have the $p$-form and the $d-p-1$-form Goldstones in the same coset construction. There are therefore dual {\it physically} equivalent descriptions. In particular, one effective action involves the $p$-form Goldstone and the other involves the $d-p-1$-form Goldstone. 
\item In non-equilibrium systems, Goldstone-like excitations exist even for unbroken $p$-form symmetries. The distinguishing feature between spontaneously broken and unbroken $p$-form Goldstones is the emergence of a  time-independent gauge symmetry. These findings are a natural extension of those in~\cite{Landry}. The conserved $d-p-1$-form charge that would exist if the $p$-form symmetry were spontaneously broken becomes gauge non-invariant and thus fails to be a physical charge. 
\item There is a second way to construct an action with an unbroken $U(1)$ $p$-form symmetry. We work in the dual picture involving the $d-p-1$-form Goldstone, but mandate that the $d-p-1$-form symmetry be a non-Stückelberg symmetry. This leads to non-conservation of the $d-p-1$-form current but does not affect the conservation of the $p$-form current. As a result, the $p$-form symmetry, as it enjoys no mixed 't Hooft anomaly, cannot be spontaneously broken. Further, we can give the $d-p-1$-form current an arbitrarily weak non-conservation, meaning that we can, in a certain sense, have a weakly {\it unbroken }$p$-form symmetry. 
\item Lastly, using the coset construction, we were able to reproduce the well-known Chern-Simons action. That is, from symmetry principles alone, we were able to construct a purely topological theory. 
\end{itemize}

This work admits generalizations and applications in many directions. First, there is an interesting consequence of these findings that warrants further investigation: in non-equilibrium systems, whether or not a given symmetry is spontaneously broken, may depend on the time-scale over which it is observed. Consider the situation of a $p$- and a $d-p-1$-form symmetry with mixed 't Hooft anomaly.  If we consider the $d-p-1$-form symmetry to be a non-Stückelberg symmetry, the strength of the non-conservation of the current is characterized by a time-scale $\tau$ in the sense that for frequency $\omega \tau \ll 1$, the current is not conserved, whereas for $\omega\tau \gg 1$, the current appears conserved. In this way, non-conservation clearly exists in the low-frequency regime, meaning that the $p$-form symmetry is unbroken. However in the high-frequency regime, the non-conservation is highly suppressed, meaning that the $p$-form symmetry will appear broken. In this paper, we saw that our quadratic models for electromagnetic systems behave in this way, but it appears to be a more general phenomenon. Second, the existing non-equilibrium effective actions for magnetohydrodynamic systems do not account for situations in which the matter is weakly charged. By exploiting principles of mixed 't Hooft anomalies and non-Stückelberg symmetries, we were able to construct a quadratic magnetohydrodynamic action that permits under-damped waves, characteristic of the small-charge limit. Further, the fact that such actions involve electric and magnetic fields directly may ultimately yield important insights into the nature of ABJ anomalies  and plasma instabilities, which appear in neutron stars. Third, higher-form symmetries are often associated with topological phases of matter. The non-equilibrium coset construction involving higher-form symmetries may therefore prove to be a powerful tool for constructing theories for topological phases beyond the Chern-Simons action. Fourth, we have only considered one kind of mixing between various $p$-form symmetries, namely we considered mixed 't Hooft anomalies. However, there are other, more complicated relationships that can exist among higher-form symmetries, like two-groups~\cite{Iqbal:2020lrt}. Extending the coset construction to account for objects such as two-groups is of significant theoretical interest. Finally, higher-form symmetries have many applications in a wide variety of areas. For example, they can be used to construct effective field theories of superfluid vortices~\cite{Horn:2015zna} and we expect they should play a similar role in describing the dynamics of defects in solids in the non-equilibrium EFT formalism~\cite{1603.04254}. Understanding such topological objects from the perspective of non-equilibrium EFT may have important applications in many areas of physics including understanding vortices in certain dark matter models~\cite{FDM}.

\bigskip
\bigskip

\noindent{\bf Acknowledgments:} I would like to thank Alberto Nicolis and Lam Hui for their wonderful mentorship and Sašo Grozdanov and Matteo Baggioli for insightful conversations. This work was partially supported by the US Department of Energy grant DE-SC011941.

\end{document}